\documentclass[letterpaper,journal]{IEEEtran}
\usepackage{amsmath,amsfonts}
\usepackage{algorithmic}
\usepackage{algorithm}
\usepackage{array}
\usepackage[caption=false,font=normalsize,labelfont=sf,textfont=sf]{subfig}
\usepackage{textcomp}
\usepackage{stfloats}
\usepackage{url}
\usepackage{verbatim}
\usepackage{graphicx}
\usepackage{cite}
\usepackage{amsmath,amssymb}
\usepackage{multirow}
\usepackage{booktabs}
\usepackage{newtxtext}
\usepackage{threeparttable}
\usepackage[table]{xcolor}
\UseRawInputEncoding
\usepackage[utf8]{inputenc}
\usepackage{algorithm,algorithmic}
\usepackage{cleveref}
\usepackage{listings}
\usepackage{xcolor}
\newcommand{\eg}{e.g., }
\definecolor{stage}{RGB}{155, 194, 230}
\newcommand{\lz}[1]{{\color{black}#1}}

\begin{document}

\title{
OneEmo: A Unified Multimodal Reasoning Model for Emotion Perception, Understanding, and Interaction }

\author{
Jiahao~Huang, Zheng~Lian~\IEEEmembership{Senior~Member,~IEEE}, Jingyi~Zhang, Zhide~Chen, Xiaojiang~Peng, Shaonan~Wang
\thanks{Jiahao~Huang is with the Fujian Normal University, Fujian, and the State Key Laboratory of Autonomous Intelligent Unmanned Systems, Tongji University, Shanghai (e-mail: qsz20241923@student.fjnu.edu.cn). 
Zheng~Lian is with the State Key Laboratory of Autonomous Intelligent Unmanned Systems, Tongji University, and the Frontiers Science Center for Intelligent Autonomous Systems, Ministry of Education (e-mail: lianzheng@tongji.edu.cn). 
Jingyi~Zhang is with the School of Psychology, Fujian Normal University (e-mail: kr1st3n0104@gmail.com). 
Zhide~Chen is with the Provincial Key Lab of Network Security \& Cryptology, Fujian Normal University, Fujian (e-mail: zhidechen@fjnu.edu.cn). 
Xiaojiang~Peng is with Shenzhen Technology University, China (e-mail: pengxiaojiang@sztu.edu.cn).
Shaonan~Wang is with the Division of Artificial Intelligence and the Humanities, and the Department of Language Science and Technology, Hong Kong Polytechnic University, China (e-mail: shaonan.wang@polyu.edu.hk). Project Leader: Zheng~Lian. Corresponding authors: Zheng~Lian, Zhide~Chen, and Shaonan~Wang.}

% Geng Tu and Shengzhe Sun contribute equally to this work.
% Yang Wu is with City University of Hong Kong. E-mail: ywu474-c@my.cityu.edu.hk.\protect\\
% Zheng Lian is with Shanghai Research Institute for Intelligent Autonomous Systems, Tongji University, Shanghai 201210, China. E-mail: lianzheng@tongji.edu.cn.\protect\\
% Erik Cambria is with the College of Computing and Data Science, Nanyang Technological University, 639798, Singapore. E-mail: cambria@ntu.edu.sg.\protect\\
% Wenjie Li is with the Department of Computing, The Hong Kong Polytechnic University, Hong Kong. E-mail: cswjli@comp.polyu.edu.hk.\protect\\
% Ruifeng Xu is with the College of Artificial Intelligence, Harbin Institute of Technology (Shenzhen), Shenzhen 518055, China; also with Shenzhen Loop Area Institute, Shenzhen, China; and Peng Cheng Laboratory, Shenzhen, China. E-mail: xuruifeng@hit.edu.cn. He is the .}
%\protect\\
%Geng Tu and Shengzhe Sun contribute equally to this work.\protect\\
%Ruifeng Xu is the corresponding author.}
}

\markboth{Journal of \LaTeX\ Class Files,~Vol.~14, No.~8, August~2021}%
{Shell \MakeLowercase{\textit{et al.}}: A Sample Article Using IEEEtran.cls for IEEE Journals}


\maketitle

\begin{abstract}
% Multimodal large language models (MLLMs) demonstrate remarkable reasoning capabilities in emotion intelligence. However, existing research predominantly focuses on developing task-specific reasoning specialists, thereby neglecting inter-task synergy and under-exploring the models' latent reasoning potential. To train a unified affective reasoning generalist capable of mastering perception, understanding, and interaction, we first introduce \textit{EmoWorld-130K}, a comprehensive dataset that distills specialized affective knowledge into explicit reasoning trajectories via a rigorous human-in-the-loop workflow.  By conducting supervised fine-tuning on this dataset, we validate that multi-task mutual promotion significantly enhances emotional intelligence. Furthermore, we propose \textit{Emo-Chord}, a novel reinforcement learning training strategy that unifies multi-task reward allocation and stabilizes the optimization process, ultimately yielding our final model, \textit{OneEmo}. Extensive experiments demonstrate that OneEmo achieves highly competitive or state-of-the-art performance across eight critical tasks (multimodal sentiment analysis, emotion recognition, open-vocabulary emotion recognition, intent recognition, humor understanding, sarcasm understanding, empathetic response generation, and emotional support conversations), all while maintaining a highly efficient parameter scale. 

Multimodal Large Language Models (MLLMs) have demonstrated remarkable capabilities in emotional intelligence. However, prevailing research predominantly focuses on task-specific specialization, often neglecting inter-task synergy and leaving latent reasoning potential underexplored. To bridge this gap, we introduce \emph{OneEmo}, a unified affective generalist capable of mastering emotion perception, comprehension, and interaction. For this purpose, we first construct \emph{EmoWorld-130K}, a comprehensive dataset that distills specialized affective knowledge into explicit reasoning trajectories via a human-in-the-loop workflow. Supervised fine-tuning on this corpus reveals significant mutual benefits derived from multi-task learning. Second, to fully unlock the latent reasoning potential, we propose \emph{Emo-Chord}, a novel reinforcement learning strategy that stabilizes optimization through unified multi-task reward allocation. Extensive experiments demonstrate that OneEmo achieves state-of-the-art performance against similarly sized baselines across most benchmarks. Notably, despite having significantly fewer parameters than commercial models, OneEmo delivers highly competitive results. This paper paves the way for more reliable and interpretable affective computing. The code is
available at \textcolor{blue}{\url{https://github.com/waHAHJIAHAO/OneEmo}}.

\end{abstract}


\section{Introduction}
% % \IEEEPARstart{A}{rtificial} Emotional Intelligence demonstrates significant effects across diverse domains, 
% \lz{\IEEEPARstart{A}{rtificial} Emotional Intelligence has demonstrated significant impacts across a wide range of domains} including educational assistive systems~\cite{zhang2025emotional}, healthcare~\cite{KumarM23}, and conversational agents~\cite{Seaborn2021}. At the forefront of this evolution, multimodal large language models (MLLMs) have substantially advanced affective computing. Recent works have explored specific affective hierarchies, ranging from foundational emotion perception and understanding~\cite{cheng2024emotion, lian2025affectgpt} to advanced interaction-level capabilities, such as generating human-like empathy~\cite{zhang2025towards, 10.1145/3746027.3762029} and orchestrating clinically-guided therapeutic dialogues~\cite{hu2025beyond}.

\IEEEPARstart{A}{ffective} computing has demonstrated significant impact across diverse domains, including educational assistive systems~\cite{zhang2025emotional}, healthcare~\cite{KumarM23}, and conversational agents~\cite{Seaborn2021}. At the forefront of this field, multimodal large language models (MLLMs) have substantially advanced the development of affective computing. Recent studies have investigated the application of MLLMs to a spectrum of emotion-related tasks, ranging from foundational emotion perception and understanding~\cite{cheng2024emotion, lian2025affectgpt} to advanced interaction-level competencies, such as the generation of human-like empathy~\cite{zhang2025towards, 10.1145/3746027.3762029}.

Early efforts have predominantly relied on supervised fine-tuning (SFT). Specifically, these approaches curate instruction-following datasets that map video inputs to emotion labels, followed by additional SFT on pre-trained MLLMs to bolster emotion understanding. Representative frameworks include AffectGPT~\cite{lian2025affectgpt} and Emotion-LLaMA~\cite{cheng2024emotion}, which prioritize multimodal fusion and emotion-specific multimodal inputs, respectively. Despite their efficacy, supervised paradigms are inherently limited in eliciting the latent reasoning potential of MLLMs. To bridge this gap, inspired by the remarkable success of reinforcement learning (RL) in reasoning tasks~\cite{deepseek-r1}, researchers are increasingly integrating RL paradigms into affective computing. Such endeavors have yielded consistent performance gains, enhancing MLLM capabilities in basic emotion recognition~\cite{zhao2025r1, Fang2026, facial_r12026, huang2026nano-emox}, fine-grained emotion perception~\cite{Lian2025affectgptr1}, and multi-turn affective dialogue~\cite{Huy_Multimood}.

However, existing approaches predominantly focus on task-specific emotion specialists, often overlooking inter-task synergy. In human-computer interaction (HCI), emotion perception, understanding, and the execution of appropriate emotion interactions are seamlessly integrated~\cite{10.1145/3706599.3706743}. Given their strong correlations, recent efforts such as Nano-EmoX~\cite{huang2026nano-emox} and VidEmo~\cite{zhang2025VidEmo} have begun to leverage this synergy. \emph{Nevertheless, current research still falls short of treating perception, understanding, and interaction as a joint optimization objective, leaving the development of a unified emotion foundation model underexplored.} Furthermore, adapting RL paradigms to such a holistic framework presents significant challenges. Current emotion reasoning datasets are largely confined to perception tasks, lacking explicit reasoning trajectories for higher-order cognitive processes (e.g., intent inference and dialogue generation). Meanwhile, integrating heterogeneous tasks via RL introduces severe optimization hurdles. Optimizing under disparate reward signals frequently triggers gradient conflicts and policy collapse, thereby destabilizing convergence and hindering the realization of a robust affective generalist.

To address these challenges, we propose \textbf{OneEmo}, a unified reasoning model explicitly designed for emotional intelligence. To achieve this, we introduce \textbf{EmoWorld-130K}, a multi-level reasoning dataset spanning from emotion perception to interaction, and \textbf{Emo-Chord}, a novel multi-task reinforcement learning strategy that mitigates model collapse. As illustrated in Fig.~\ref{fig:emoworld}, \emph{EmoWorld-130K} encompasses eight core affective tasks: Multimodal Sentiment Analysis (MSA), Basic Emotion Recognition (B-MER), Open-Vocabulary MER (OV-MER), Intent Recognition (MIR), Humor Understanding (MHU), Sarcasm Understanding (MSU), Empathetic Response Generation (ERG), and Emotional Support Conversation (ESC). Each instance is annotated with theory-driven, structured reasoning trajectories and corresponding answers. To overcome multi-task optimization instability, we propose \emph{Emo-Chord}, a novel RL strategy that employs an off-policy cold start followed by a hybrid optimization phase combining Group Relative Policy Optimization (GRPO)~\cite{2024DeepSeekMath} with a dynamically weighted SFT auxiliary loss. This design enables continuous replay of expert data, preventing capability degradation during exploration. Guided by a synergistic rubric-based reward system, Emo-Chord enforces reasoning coherence and multimodal factual grounding. Extensive experiments show that OneEmo achieves state-of-the-art results against similarly sized baselines across most benchmarks. Our main contributions are as follows:

\begin{figure*}[t]
\centering
\includegraphics[width=1.0\textwidth]{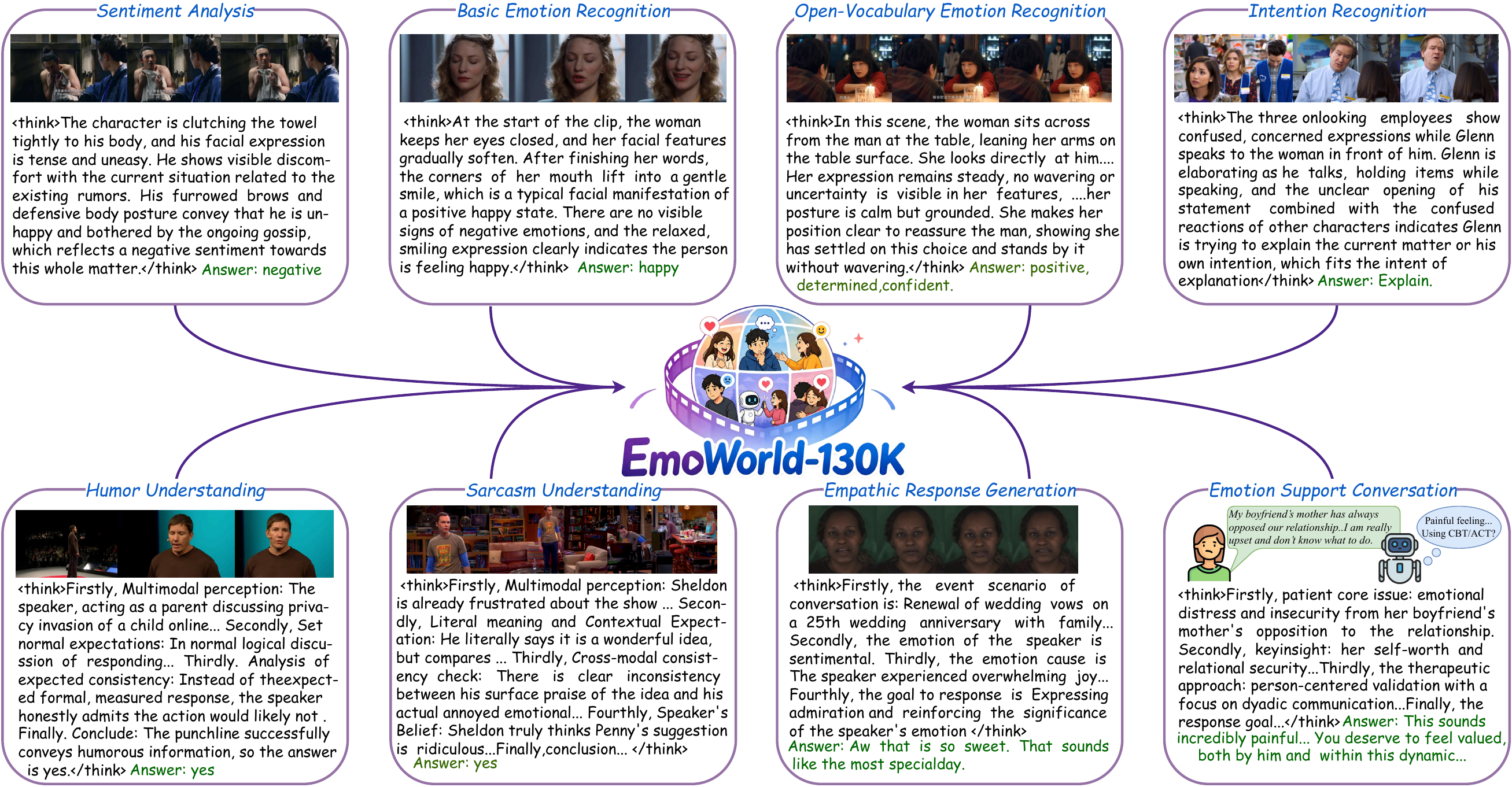}
\caption{% Ours EmoWorld-130K dataset provides comprehensive multi-level emotion task reasoning trajectories. It systematically covers a broad spectrum of capabilities, advancing from emotion perception (sentiment analysis and emotion recognition) and understanding (conversational styles and social intent comprehension) toward anthropomorphic interaction (empathetic response and emotional support).
\textbf{EmoWorld-130K Dataset}. This dataset provides high-quality reasoning trajectories for eight affective tasks encompassing three domains: emotion perception (sentiment analysis, basic emotion recognition, and fine-grained emotion recognition), emotion understanding (intent recognition, humor detection, and sarcasm detection), and emotion interaction (empathetic response generation and emotional support conversation).
}
\label{fig:emoworld}
\end{figure*}

\begin{itemize}
    \item We present \textbf{OneEmo}, a unified multimodal reasoning framework capable of handling eight critical affective tasks, spanning foundational emotion perception to advanced multi-turn emotion interactions.
    
    \item To build {OneEmo}, we first introduce \textbf{EmoWorld-130K}, a comprehensive dataset featuring psychology-informed reasoning trajectories. Meanwhile, we propose \textbf{Emo-Chord}, a unified multi-task RL strategy that integrates on-policy exploration with off-policy imitation. By using fine-grained reasoning rewards, Emo-Chord ensures high-quality inference while stabilizing cross-task synergy.
    
    \item Extensive automatic and human evaluations demonstrate that {OneEmo} achieves highly competitive benchmark results and superior interaction capabilities.
\end{itemize}

%%%%%%%%%%%%%%%%%%%%%%%%%%%%%%%%%%%%%%fig3%%%%%%%%%%%%%%%%%%%%%%%%%%%%%%%%%%%%%%%%%%%%%%%%%%%%%%
\begin{figure*}[t]
\centering
\includegraphics[width=1.0\textwidth]{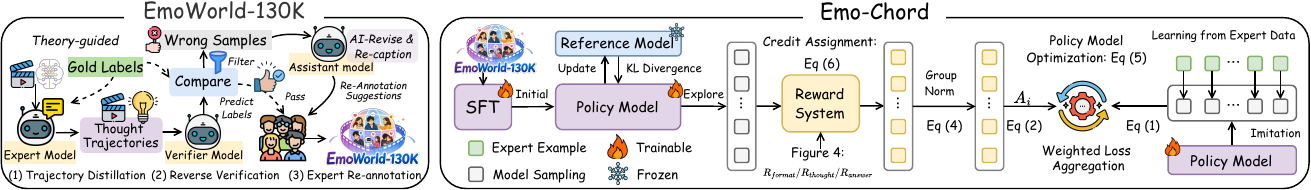}
\caption{
% \textbf{The overarching framework of our proposed approach}. \textbf{(Left)} The data annotation workflow, encompassing reasoning trajectory distillation from expert models, automated answer verification via reverse reasoning, and manual audit by human experts. \textbf{(Right)} The Emo-Chord training strategy, which optimizes the model using RL augmented with an auxiliary SFT loss, immediately following an SFT cold start, which will more effectively stimulate the multi-task reasoning potential of OneEmo. 
\textbf{OneEmo} utilizes EmoWorld-130K as the training corpus and adopts the Emo-Chord training strategy. \textbf{EmoWorld-130K (Left)} is constructed via a three-stage pipeline: reasoning trajectory distillation from expert models, automated answer verification via reverse reasoning, and rigorous manual checks.  \textbf{Emo-Chord (Right)} initiates with SFT, followed by RL optimization augmented with an auxiliary SFT loss, eliciting the multi-task reasoning capabilities.
}
\label{fig:framework}
\end{figure*}
%%%%%%%%%%%%%%%%%%%%%%%%%%%%%%%%%%%%%%%fig3%%%%%%%%%%%%%%%%%%%%%%%%%%%%%%%%%%%%%%%%%%%%%%%%%%%%%

\section{Related Works}
% \textbf{Multimodal Emotion Language Models (MELMs)} 
% typically consist of a multimodal encoder, an adapter, and a large language model. Compared to traditional architectures, MELMs demonstrate profound advantages in affective computing, driven by their superior representation perception and causal reasoning capabilities. Recently, researchers~\cite{cheng2024emotion, lian2025affectgpt, huang2025emotion, yang2025omni, 11347465} achieved explainable emotion recognition by leveraging supervised fine-tuning. Furthermore, Zhou et al.~\cite{zhou2026HIER} employ structured reasoning alongside a feedback-driven self-evolution module to achieve fine-grained intent comprehension. Similarly, Schmidmaier et al.~\cite{10.1145/3701716.3718371} facilitate intent recognition and multi-turn interactions through hierarchical semantics and adaptive reasoning, while E3RG~\cite{10.1145/3746027.3762029} enables training-free empathetic interactions via prompt engineering and empathetic memory retrieval.
% However, these existing approaches predominantly focus on optimizing performance for specific tasks, inherently neglecting the potential for knowledge transfer and synergy across diverse affective tasks. To this end, OneEmo synergistically optimizes eight core tasks. This allows the model to comprehensively assimilate emotional domain knowledge, ultimately demonstrating immense potential for real-world human-computer interaction.

\subsection{Multimodal Emotion Language Models (MELMs)}

MELMs typically comprise a multimodal encoder, an adapter, and a large language model. Benefiting from superior perceptual and reasoning capabilities, MELMs have demonstrated distinct advantages over traditional architectures. Recent advances have leveraged supervised fine-tuning to enable explainable emotion recognition~\cite{cheng2024emotion, lian2025affectgpt, huang2025emotion, yang2025omni, 11347465}. Building on this, Zhou et al.~\cite{zhou2026HIER} integrate structured reasoning with a feedback-driven self-evolution module to achieve fine-grained intent comprehension. Schmidmaier et al.~\cite{10.1145/3701716.3718371} facilitate intent recognition and multi-turn interaction via hierarchical semantic modeling and adaptive reasoning. Lin et al.~\cite{10.1145/3746027.3762029} propose a training-free paradigm for empathetic response generation utilizing prompt engineering and empathetic memory retrieval. Despite these successes, current approaches predominantly target task-specific optimization. This focus overlooks the inherent synergies across diverse affective tasks, even though emotion perception, understanding, and interaction are intrinsically interconnected \cite{10.1145/3706599.3706743}. To bridge this gap, we introduce OneEmo, a unified multimodal reasoning model designed to comprehensively assimilate emotional domain knowledge.

% \textbf{Reinforcement Learning in MELMs:}
% Recently, RL post-training has emerged as a powerful paradigm to unlock the reasoning potential of MELMs, providing causal interpretability through explicit thought trajectories. Within perception-level tasks, R1-Omni~\cite{zhao2025r1} and AffectGPT-R1~\cite{Lian2025affectgptr1} pioneer the use of verifiable rewards for the B-MER task, while ERV~\cite{Rha_Yeo_Kim_Ro_2026} introduces a rationale verifier to rectify reasoning-response inconsistencies. Concurrently, for interaction-level tasks, approaches like EmpRL~\cite{10899840} and MultiMood~\cite{Huy_Multimood} optimize generation quality utilizing empathy, credibility, and similarity metrics as reward signals.
% Despite their success in basic perception, these approaches exhibit limited interpretability when scaling to complex understanding, interactive tasks. We leverage the large-scale EmoWorld-130K dataset alongside the GRPO~\cite{2024DeepSeekMath} algorithm for emotion multi-task RL. This paradigm significantly enhances both the interpretability and generalization capabilities of MELMs.

\subsection{Reinforcement Learning in MELMs}

RL has emerged as a pivotal paradigm for eliciting the reasoning capabilities of MELMs. Pioneering works such as R1-Omni~\cite{zhao2025r1} and AffectGPT-R1~\cite{Lian2025affectgptr1} leverage RL to enhance emotion perception. Subsequent studies, including ERV~\cite{Rha_Yeo_Kim_Ro_2026}, introduce rationale verifiers to mitigate reasoning-response inconsistencies. Furthermore, EmpRL~\cite{10899840} and MultiMood~\cite{Huy_Multimood} refine generative quality by utilizing empathy, credibility, and semantic similarity as reward signals. Despite these advances, extending RL to jointly optimize emotion perception, understanding, and interaction remains challenging. The integration of such heterogeneous tasks often leads to unstable optimization dynamics and conflicting gradients. To address this, we introduce the EmoWorld-130K dataset and the Emo-Chord optimization strategy. Together, they enhance the performance and interpretability of MELMs.

\section{Preliminary}

Current post-training paradigms for MLLMs primarily comprise SFT and RL. Specifically, SFT utilizes curated, domain-specific instruction–response pairs. By learning to mimic the target response distribution, the model minimizes the token-level negative log-likelihood, which is formulated as:
\begin{equation}
\mathcal{L}_{\text{SFT}}(\theta) = -\mathbb{E}_{(x, y) \sim \mathcal{D}} \left[ \sum_{t=1}^{|y|} \log \pi_{\theta}(y_t \mid x, y_{<t}) \right],
\label{eq:sft}
\end{equation}
where $x$  denotes the multimodal input context and $y$ represents the target token sequence.

RL optimizes a policy model using feedback signals from a reward-based credit assignment system. This paper adopts GRPO~\cite{2024DeepSeekMath}, which estimates advantages by normalizing rewards across $G$ sampled outputs for each input:
\begin{equation}
\begin{split}
\mathcal{L}_{\text{GRPO}}(\theta) &= -\mathbb{E}_{(x, \{y_i\}_{i=1}^{G}) \sim \mathcal{D}} \Bigg[ \frac{1}{G} \sum_{i=1}^{G} \frac{1}{|y_i|} \sum_{t=1}^{|y_i|} \\
&\qquad \min\Bigl( r_{i,t} \hat{A}_i,\; \text{clip}\bigl(r_{i,t}, 1-\epsilon, 1+\epsilon\bigr) \hat{A}_i \Bigr) \Bigg] \\
&\quad + \beta\, \mathbb{D}_{\text{KL}}\bigl(\pi_{\theta} \| \pi_{\text{ref}}\bigr).
\end{split}
\label{eq:grpo}
\end{equation}
Here, a KL divergence penalty regularizes the updated policy $\pi_{\theta}$ toward the reference policy $\pi_{\text{ref}}$, mitigating policy drift during optimization. The probability ratio $r_{i,t}$ is computed via importance sampling between $\pi_{\theta}$ and $\pi_{\text{ref}}$:
\begin{equation}
r_{i,t} = \frac{\pi_{\theta}(y_{i,t} \mid x, y_{i,<t})}{\pi_{\theta_{\text{old}}}(y_{i,t} \mid x, y_{i,<t})}.
\label{eq:importance_sampling}
\end{equation}
The intra-group relative advantage $\hat{A}_i$ is derived by normalizing the raw reward $R_i$ against the mean and standard deviation of the reward group:
\begin{equation}
\hat{A}_i = \frac{R_i - \text{mean}\bigl(\{R_j\}_{j=1}^{G}\bigr)}{\text{std}\bigl(\{R_j\}_{j=1}^{G}\bigr)}.
\label{eq:advantage}
\end{equation}

\section{Methodology}

Fig.~\ref{fig:framework} presents the overall pipeline of OneEmo, which takes EmoWorld-130K as the training corpus and Emo-Chord as the optimization strategy. This section elaborates on the dataset construction process and the proposed training pipeline.

\subsection{EmoWorld-130K: Theory-driven Reasoning Dataset}

% Conventional emotion datasets, \eg ECR-Chain~\cite{ijcai2024p0695}, Psyche-R1~\cite{dai2025psycher1reliablepsychologicalllms}, and Emo-CFG~\cite{zhang2025VidEmo}, predominantly target isolated affective dimensions, restricting their focus to either low-level perception or high-level interaction. Such a fragmented paradigm inherently neglects the synergistic relationships across diverse tasks and falls short in providing comprehensive interpretability for complex emotional dynamics. To overcome these limitations and cultivate unified multi-task affective reasoning capabilities, we introduce the \textit{EmoWorld-130K} dataset, the statistical distribution and detailed composition of the dataset are summarized in Figure~\ref{fig:emoworld-proportion}. As depicted in the left panel of Figure~\ref{fig:framework}, the dataset is constructed through a rigorous three-step curation workflow:

Conventional emotion datasets, \eg ECR-Chain~\cite{ijcai2024p0695}, Psyche-R1~\cite{dai2025psycher1reliablepsychologicalllms}, and Emo-CFG~\cite{zhang2025VidEmo}, predominantly target isolated affective dimensions. This fragmented paradigm neglects the synergistic relationships across tasks and fails to provide comprehensive interpretability. Moreover, these datasets leave the latent reasoning potential of MLLMs underexplored. To bridge this gap, we introduce \textit{EmoWorld-130K}, which is constructed through a three-step workflow, as depicted in Fig.~\ref{fig:framework} (Left). Statistical distribution and detailed composition of the dataset are summarized in Fig.~\ref{fig:emoworld-proportion}.

\subsubsection{Theory-based Trajectory Distillation} 

We first source raw data from established datasets, including DFEW~\cite{jiang2020dfew}, MERR~\cite{cheng2024emotion}, MER-Caption+~\cite{lian2025affectgpt}, MER2025-OV~\cite{lian2025mer}, MIntRec~1.0/2.0~\cite{MIntRec,zhang2024mintrec}, Mustard~\cite{mustard}, URFunny~\cite{hasan-etal-2019-ur}, AvaMERG~\cite{zhang2025towards}, and Openr1\mbox{-}Psy~\cite{hu2025beyond}. Using original annotations as label anchors, we prompt Seed-2.0-Lite~\cite{bytedance2026seed2} to generate structured reasoning trajectories. \emph{Rather than allowing unconstrained generation, we operationalize well-established psychological and cognitive frameworks into explicit prompt templates to rigorously govern the model's reasoning logic.} Specifically, for basic emotion perception, the model extracts multimodal cues and deduces labels based on Ekman's Theory~\cite{ekman1992argument}, which compels the LLM to anchor its inferences on observable facial expressions and behavioral markers rather than abstract guesses. For humor and sarcasm understanding, it analyzes cognitive expectation mismatches and deciphers pretended attitudes, grounded in the Incongruity-Resolution Theory~\cite{suls1972twostage}, the Psychology of Humour~\cite{goldstein1972psychology}, and the Theory of Irony~\cite{clark1984pretense}. For emotion interaction, driven by the Appraisal Theory of Emotion~\cite{lazarus1991emotion}, we formulate a four-step reasoning paradigm: dialog memory tracking, user state perception, causal attribution, and response goal formulation. Notably, the emotional support conversations task additionally integrates DSM/ICD\mbox{-}11 diagnostic criteria~\cite{apa2013dsm5, who2022icd11} to standardize clinical state evaluation, and therapeutic strategies (e.g., CBT~\cite{beck1976cognitive}, ACT~\cite{hayes1999act}) to logically dictate the selection of supportive techniques, ensuring highly explainable responses. Compared to the unstructured and verbose thought in OpenR1-Psy~\cite{hu2025beyond}, our generated trajectories remain highly structured, token-efficient, and strictly context-aware.

\subsubsection{Automated Reverse Verification}

To ensure factual consistency and logical coherence in thought annotations, we introduce a closed-loop validation protocol. This protocol assesses whether the gold labels can be deduced solely from the distilled reasoning trajectories. Samples exhibiting inconsistencies are flagged with specific violation types and causal rationales, which are then prioritized for human annotation.

\subsubsection{AI-assisted Human Expert Re-annotation} 

To address the identified inconsistencies, three psychology postgraduates meticulously reviewed and revised the problematic trajectories, incorporating targeted repair suggestions generated by AI. To establish rigorous quality control, we manually audited a random 10\% subset of the data for each task. If any subset failed to meet our stringent quality criteria, we initiated iterative re-annotation and repeated sampling inspections until the entire dataset satisfied our high-quality standards.

%%%%%%%%%%%%%%%%%%%%%%%%%%%%%%%%%%%%%%fig2%%%%%%%%%%%%%%%%%%%%%%%%%%%%%%%%%%%%%%%%%%%%%%%%%%%%%%
\begin{figure}[t]
\centering
\includegraphics[width=\columnwidth]{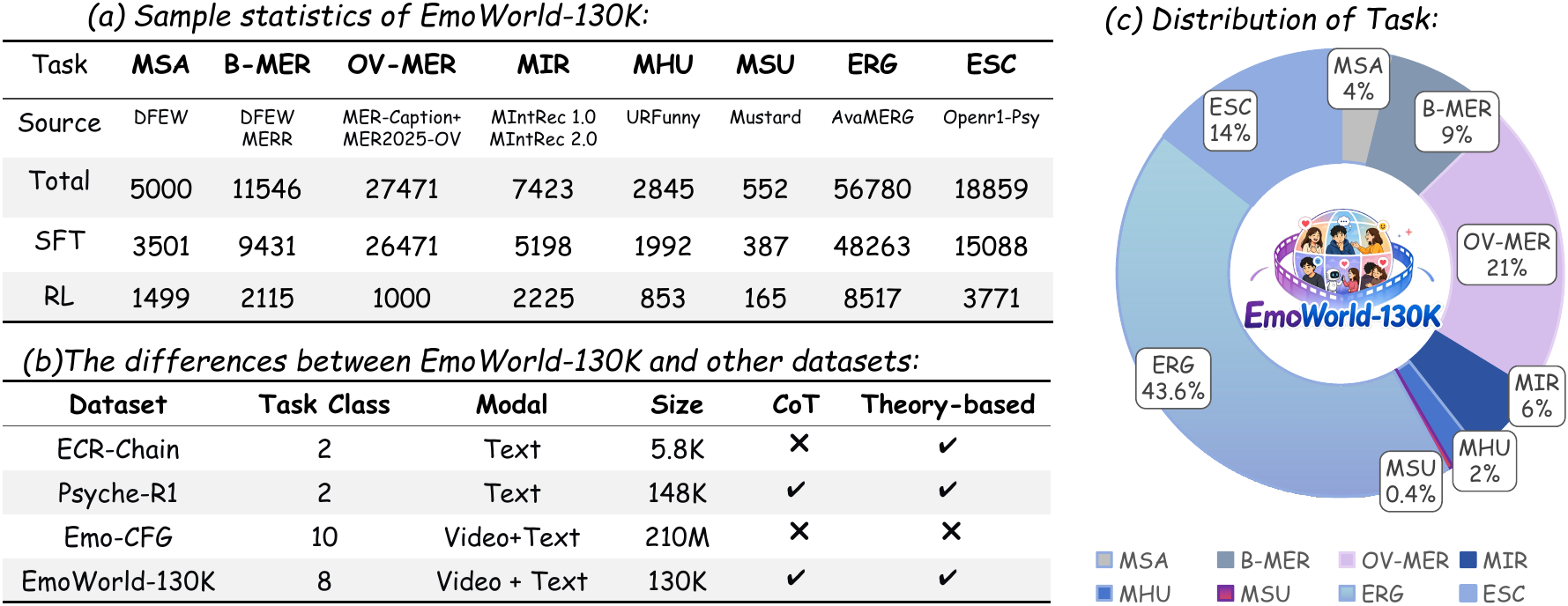}
\caption{Statistics of EmoWorld-130K and comparison with other datasets.}
\label{fig:emoworld-proportion}
\end{figure}
%%%%%%%%%%%%%%%%%%%%%%%%%%%%%%%%%%%%%%%fig2%%%%%%%%%%%%%%%%%%%%%%%%%%%%%%%%%%%%%%%%%%%%%%%%%%%%%

\subsection{Emo-Chord: Multi-task Collaborative Optimization}

Emo-Chord comprises two core modules: training strategy and credit assignment. Additionally, we introduce a Task-Aware Linear Decay Gating mechanism to regulate reasoning length and mitigate potential hallucinations.

\subsubsection{Training Strategy}

To validate that synergistic learning across multi-level affective tasks optimizes overall performance, we first employ a two-stage curriculum learning on the EmoWorld-130K SFT subset (Fig.~\ref{fig:emoworld-proportion}). In the first stage, the model undergoes joint fine-tuning on diverse multimodal tasks to acquire emotion perception and understanding capabilities. In the second stage, we introduce the ESC task and substantially increase the proportion of ERG data. This compels the model to analyze multi-turn dialogues, historical emotional trajectories, and interaction dynamics to generate supportive responses, thereby bridging perception with interaction.

While RL is adept at eliciting reasoning capabilities, we observe that models are susceptible to policy collapse and reward instability, creating significant bottlenecks for RL optimization. Drawing inspiration from recent advances~\cite{MIXCHORD}, we mitigate this by integrating an auxiliary SFT loss during training:
\begin{equation}
\mathcal{L}(\theta) = (1-\mu) \mathcal{L}_{\text{GRPO}}(\theta) + \mu \mathcal{L}_{\text{SFT}}(\theta),
\label{eq:training_object}
\end{equation}
where $\mu$ denotes a dynamic balancing coefficient that controls the relative contribution of each loss component.

% Fig.~\ref{fig:framework} (right) illustrates the Emo-Chord training pipeline. To prevent sub-optimal policy entrapment while sustaining open-ended exploration, we initiate an offline cold-start using Stage 1 of Curriculum SFT, followed by on-policy GRPO on the \textit{EmoWorld-130K} RL subset (Fig.~\ref{fig:emoworld-proportion}).  This progressive paradigm effectively cultivates multi-task reasoning capabilities while substantially enhancing training stability and downstream performance. We systematically validate this strategy against alternative pipelines in Section~\ref{sec:experiment_training_strategy}.

Fig.~\ref{fig:framework} (right) illustrates the overall training pipeline. To circumvent sub-optimal policy entrapment while preserving exploratory diversity, we initiate training with an offline cold-start phase, followed by online policy optimization on \textit{EmoWorld-130K}. This hybrid paradigm effectively cultivates robust multi-task reasoning capabilities while substantially enhancing both training stability and final model performance. We systematically compare this strategy against alternative pipelines in Section~\ref{sec:experiment_training_strategy}.

%%%%%%%%%%%%%%%%%%%%%%%%%%%%fig4%%%%%%%%%%%%%%%%%%%%%%%%%%%%%%%%%%%%%%%%
\begin{figure*}[t]
\centering
\includegraphics[width=1.0\textwidth]{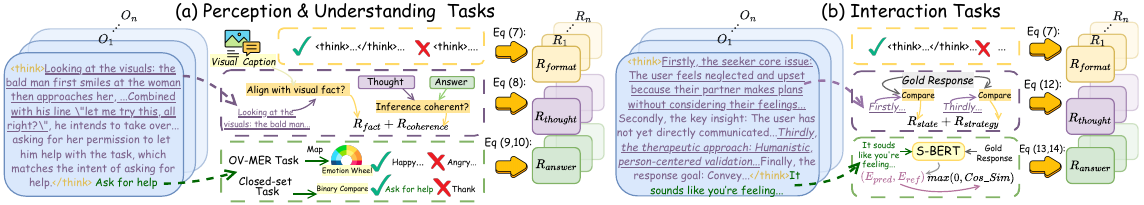}
\caption{
% \textbf{Illustration of the Emo-Chord credit assignment mechanism.}. The total reward integrates format, thought, and answer components, balanced by coefficients $\gamma_{i}$. For perception and understanding tasks, the framework rigorously aligns factual details and reasoning coherence, calculating answer rewards directly. For interactions tasks, it models user states and aligns response strategies, optimizing the final reward based on target response similarity. 
\textbf{Credit assignment in Emo-Chord}. Emo-Chord decomposes rewards into format, thought, and answer components across three domains (perception, understanding, and interaction). For perception and understanding, we enforce constraints on visual details and logical consistency. For interaction, we align the model’s inferred user states and response strategies with reference gold responses, refining the final reward based on semantic similarity.}
\label{fig:credit_assignment}
\end{figure*}

\subsubsection{Credit Assignment}
Fig.~\ref{fig:credit_assignment} details the reward functions used in Emo-Chord. For emotion perception, understanding, and interaction, these functions comprise three key components: format, thought, and answer. They are aggregated via coefficients $\gamma_{f}$, $\gamma_{t}$, and $\gamma_{a}$:
\begin{equation}
R = \gamma_{f} R_{\text{format}} + \gamma_{t} R_{\text{thought}} + \gamma_{a} R_{\text{answer}}.
\label{eq:total_reward}
\end{equation}

$R_{\text{format}}$ applies across all tasks as a binary reward enforcing structural compliance: the reasoning process must be strictly encapsulated within \texttt{<think>} and \texttt{</think>} tags, immediately followed by the final answer:
\begin{equation}
R_{\text{format}}(y) = 
\begin{cases}
1, & \text{if } y \in \mathcal{F} \\
0, & \text{otherwise}
\end{cases}.
\label{eq:format_reward}
\end{equation}

$R_{\text{thought}}$ and $R_{\text{answer}}$ are task-dependent, evaluating the quality of the reasoning trajectory and the final response, to ensure both interpretability and accuracy. We tailor these rewards across two distinct task categories: \textit{(a) Perception and Understanding}: The model infers specific emotional states, dialogue styles, and latent intents from multimodal cues. \textit{(b) Interaction}: The objective shifts to modeling user states, performing causal analysis, and generating empathetic or supportive responses.

% \textbf{Perception\&Understanding.} The thought reward $R_\text{thought}$ comprises factual consistency $R_\text{fact}$ and reasoning coherence $R_{\text{coherence}}$, formalized as:
% \begin{equation}
% R_{\text{thought}} = \frac{R_{\text{fact}} + R_{\text{coherence}}}{10}.
% \label{eq:tought_reward_1}
% \end{equation}
% To evaluate these components systematically, we devise detailed rubrics for an LLM judge. To compute $R_{\text{fact}}$, a video understanding expert (Qwen3.5-9B) first processes the raw video to extract atomic visual fact descriptions. The LLM then compares the visual descriptions generated in the sampled thought process against these objective facts, assigning a discrete score from 1 to 5. This rigorously prevents factual hallucinations and fabricated details from compromising response accuracy. Similarly, for $R_{\text{coherence}}$, the LLM evaluates whether the semantic trajectory of the reasoning logically aligns with the final answer, ensuring the thought process deduces the outcome naturally without logical leaps.

\textbf{Perception \& Understanding.} $R_\text{thought}$ comprises factual consistency $R_\text{fact}$ and reasoning coherence $R_{\text{coherence}}$:
\begin{equation}
R_{\text{thought}} = \frac{R_{\text{fact}} + R_{\text{coherence}}}{10}.
\label{eq:tought_reward_1}
\end{equation}
To evaluate these components, we devise detailed rubrics for an LLM judge. To compute $R_{\text{fact}}$, a video understanding model first processes the raw video to extract atomic visual fact descriptions. The LLM then compares the visual descriptions generated in the sampled thought process against these objective facts, assigning a discrete score from 1 to 5. This rigorously prevents factual hallucinations and fabricated details from compromising response accuracy. For $R_{\text{coherence}}$, the LLM evaluates whether the semantic trajectory of the reasoning logically aligns with the final answer, ensuring the thought process deduces the outcome naturally without logical leaps.

% Regarding the $R_{\text{answer}}$, we apply a binary reward for tasks with closed-set label spaces, expressed as:
% \begin{equation}
% R_{\text{answer}}(y, o^a_i) = \delta_{y, o^a_i} = 
% \begin{cases}
% 1, & y = o^a_i \\
% 0, & y \neq o^a_i
% \end{cases}.
% \label{eq:answer_1}
% \end{equation}
% For open-vocabulary emotion recognition task, inspired by AffectGPT-R1~\cite{Lian2025affectgptr1}, we normalize the predictions via an emotion wheel and compute the Weighted Average F1 (WAF) score across five emotion dimensions, expressed as:
% \begin{equation}
% R_{\text{answer}}(y, o^a_i) = \text{WAF}(y, o^a_i) = \frac{1}{C} \sum_{c=1}^{C} w_c \cdot F_c(y, o^a_i),
% \label{eq:waf_reward}
% \end{equation}

% \begin{equation}
% F_c(y, o^a_i) = 2 \cdot \frac{\text{Precision}_c(y, o^a_i) \cdot \text{Recall}_c(y, o^a_i)}{\text{Precision}_c(y, o^a_i) + \text{Recall}_c(y, o^a_i)}.
% \label{eq:fscore}
% \end{equation}

For $R_{\text{answer}}$, we apply a binary reward for tasks with closed-set label spaces:
\begin{equation}
R_{\text{answer}}(y, o^a_i) = 
\begin{cases}
1, & y = o^a_i \\
0, & y \neq o^a_i
\end{cases}.
\label{eq:answer_1}
\end{equation}
For OV-MER, following established protocols~\cite{lian2025affectgpt, Lian2025affectgptr1}, we compute the reward based on Emotion Wheel (EW) metrics:
\begin{equation}
F_c(y, o^a_i) = 2 \cdot \frac{\text{Precision}_c(y, o^a_i) \cdot \text{Recall}_c(y, o^a_i)}{\text{Precision}_c(y, o^a_i) + \text{Recall}_c(y, o^a_i)}.
\label{eq:fscore}
\end{equation}
\begin{equation}
R_{\text{answer}}(y, o^a_i) = \frac{1}{C} \sum_{c=1}^{C} F_c(y, o^a_i),
\label{eq:waf_reward}
\end{equation}
where $C$ denotes the number of emotion wheels.

% \textbf{Interaction.}
% The thought reward for interactive tasks consists of a user state modeling reward $R_\text{{state}}$ and a strategy alignment reward $R_{\text{strategy}}$:
% \begin{equation}
% R_{\text{thought}} = \frac{R_{\text{state}} + R_{\text{strategy}}}{10},
% \label{eq:tought_reward_2}
% \end{equation}       
% $R_{\text{state}}$ evaluates the modeling accuracy (scored 1-5 by the LLM) by comparing the user state depicted in the sampled thought $o^t_i$ against the ground-truth response $y^t_{gt}$. Accurately grasping the user's current situation, emotional state, core conflicts, and underlying drivers is critical for formulating effective strategies. Similarly, $R_{\text{strategy}}$ assesses the strategic alignment between $o^t_i$ and $y^t_{gt}$. The LLM assigns higher scores (1-5) to generated strategies that are identical or broadly consistent with the ground-truth.
% For $R_{\text{answer}}$ in these open-ended tasks, we extract embedding pairs $(E_{pred},E_{ref})$ for the sampled answer $o^a_i$ and the ground-truth answer $y^a_{gt}$ using Sentence-BERT (S-BERT)~\cite{reimers-gurevych-2019-sentence} which serves as deliberately weak feedback signal: it anchors responses to the reference distribution without enforcing lexical overlap. This process is formalized as follows:
% \begin{equation}
% E_{pred} = \text{S-BERT}(o^{a}_{i}), \quad E_{ref} = \text{S-BERT}(y^{a}_{gt}),
% \label{eq:embedding_pair}
% \end{equation}

% \begin{equation}
% R_{\text{answer}} = \max(0, \text{Similarity}(E_{\text{pred}}, E_{\text{ref}})).
% \label{eq:answer_2}
% \end{equation}

\textbf{Interaction.}
$R_{\text{thought}}$ comprises a user state modeling reward $R_\text{{state}}$ and a strategy alignment reward $R_{\text{strategy}}$. Accurately capturing the user’s situational context, emotional state, and underlying motivations is critical for response generation. $R_{\text{state}}$ evaluates the fidelity of user modeling (scored 1–5 by an LLM judge) by comparing the inferred state against the gold reference. Similarly, $R_{\text{strategy}}$ assesses the alignment of the proposed intervention strategy with the gold reference.
\begin{equation}
R_{\text{thought}} = \frac{R_{\text{state}} + R_{\text{strategy}}}{10}.
\label{eq:tought_reward_2}
\end{equation}

For $R_{\text{answer}}$ in these open-ended tasks, we embed the predicted response and the gold response using Sentence-BERT (S-BERT)~\cite{reimers-gurevych-2019-sentence} and calculate their similarity:
\begin{equation}
E_{pred} = \text{S-BERT}(o^{a}_{i}), \quad E_{ref} = \text{S-BERT}(y^{a}_{gt}),
\label{eq:embedding_pair}
\end{equation}
\begin{equation}
R_{\text{answer}} = \max\left(0, \text{Similarity}(E_{\text{pred}}, E_{\text{ref}})\right).
\label{eq:answer_2}
\end{equation}

% \subsubsection{Task-Aware Linear Decay Gating}
% Affective tasks require heterogeneous reasoning lengths. During the exploration process, the model  overgeneralize verbose reasoning to simpler tasks, inducing task-irrelevant hallucinations and degrading performance. To mitigate this, we introduce a task-aware linear decay factor $\tau$, to penalize the base answer reward $R_{\text{answer}}^{*}$. Specifically, we define task-specific soft and hard length limits. For a generated sequence of length $l_{o}$, $\tau$ is formalized as:
% \begin{equation}
% \tau =
% \begin{cases}
% 1 & l_{\text{o}} \leq l_{\text{soft}} \\ 
% \frac{l_{\text{hard}}-l_{\text{o}}}{l_{\text{hard}}-l_{\text{soft}}} & l_{\text{soft}} < l_{o} < l_{\text{hard}} \\
% 0 & l_{\text{o}} \geq l_{\text{hard}}\\
% \end{cases}.
% \label{eq:tau}
% \end{equation}
% We compute independent decay factors for the reasoning trajectory $\tau_{\text{thought}}$ and the final response $\tau_{\text{answer}}$. The final answer reward is then modulated by the minimum of these two factors:
% \begin{equation}
% R_{\text{answer}} = R_{\text{answer}}^{*} \times \min(\tau_{\text{thought}}, \tau_{\text{answer}}).
% \label{eq:linear_decay_answer_reward}
% \end{equation}
% Through this fine-grained gating mechanism, the model adapts its reasoning and generation granularity to specific task requirements, fundamentally preventing unbounded verbosity and inefficient reasoning. Please see appendix for the statistical determination of these bounds.

\subsubsection{Task-Aware Linear Decay Gating}

Emotion tasks exhibit heterogeneous complexity, necessitating variable chain-of-thought lengths. Empirically, we observe that unconstrained models over-generalize verbose reasoning and answers to simple tasks, inducing task-irrelevant hallucinations and degrading performance. To mitigate this, we introduce a task-aware linear decay factor $\tau$ to regularize the answer reward $R_{\text{answer}}$. Specifically, we define task-specific soft and hard length bounds. For a generated sequence of length $l_{o}$, $\tau$ is defined as:
\begin{equation}
\tau =
\begin{cases}
1 & l_{\text{o}} \leq l_{\text{soft}} \\ 
\frac{l_{\text{hard}}-l_{\text{o}}}{l_{\text{hard}}-l_{\text{soft}}} & l_{\text{soft}} < l_{o} < l_{\text{hard}} \\
0 & l_{\text{o}} \geq l_{\text{hard}}\\
\end{cases}.
\label{eq:tau}
\end{equation}
We compute decay factors for the reasoning trajectory and the final answer. $R_{\text{answer}}$ is then modulated by the minimum of these factors to enforce brevity across both components:
\begin{equation}
R_{\text{answer}} = R_{\text{answer}} \times \min(\tau_{\text{thought}}, \tau_{\text{answer}}).
\label{eq:linear_decay_answer_reward}
\end{equation}
This mechanism adaptively calibrates the answer granularity to specific task requirements, effectively suppressing unbounded verbosity and inefficient reasoning. The statistical procedure for determining these bounds is detailed in the Appendix.

%-------------------------------benchmarks tab-----------------------------------------
\begin{table}[!t]
% \small
\centering
\caption{\textbf{Evaluation benchmark.} 
%We train our model on EmoWorld-130K. 
To prevent data leakage, we ensure strict partition protocols. For perception, training and testing samples are sourced from disjoint datasets; for understanding and interaction, we adhere to the official dataset splits,  employing the training and validation sets for training and the test splits for final evaluation.}
\label{tab:benchmarks}
\resizebox{\columnwidth}{!}{%
\begin{tabular}{l|l|l}
\toprule
\textbf{Task} & \textbf{EmoWorld-130K (Training)} & \textbf{Testing} \\
\midrule
\multicolumn{3}{c}{\textit{Emotion Perception}} \\
\midrule
MSA    & DFEW~\cite{jiang2020dfew} &  CMU-MOSI/MOSEI~\cite{zadeh2016mosi, zadeh2018multimodal}, CH-SIMS v1/v2~\cite{yu2020ch,liu2022make} \\
B-MER  & DFEW~\cite{jiang2020dfew}, MERR~\cite{cheng2024emotion} & MER2023/2024~\cite{lian2023mer, lian2024mer}, MELD~\cite{poria-etal-2019-meld}, IEMOCAP~\cite{busso2008iemocap} \\
OV-MER & MER-Caption+~\cite{lian2025affectgpt}, MER2025-OV~\cite{lian2025mer} & OV-MERD~\cite{lian2025ov} \\
\midrule
\multicolumn{3}{c}{\textit{Emotion Understanding}} \\
\midrule
MSU    & MUSTARD~\cite{mustard} (Train/Val) &  MUSTARD~\cite{mustard} (Test)  \\
MHU    & URFunny~\cite{hasan-etal-2019-ur} (Train/Val) & URFunny~\cite{hasan-etal-2019-ur}  (Test) \\
MIR    & MIntRec1.0/2.0~\cite{MIntRec, zhang2024mintrec} (Train/Val) & MIntRec1.0/2.0~\cite{MIntRec, zhang2024mintrec} (Test)\\
\midrule
\multicolumn{3}{c}{\textit{Emotion Interaction}} \\
\midrule
ERG    & AvaMERG~\cite{zhang2025towards} (Train/Val) & AvaMERG~\cite{zhang2025towards}  (Test) \\
ESC    & OpenR1-Psy~\cite{hu2025beyond} (Train/Val) & OpenR1-Psy~\cite{hu2025beyond} (Test) \\
\bottomrule
\end{tabular}
}
\end{table}

%------------------------------------------------------------------------------

%---------------------------------main result 6 tasks---------------------------------
\begin{table*}[t]
\centering
%\footnotesize
\scriptsize
\setlength{\tabcolsep}{1.1pt}
\lz{
\caption{\textbf{Main results on emotion perception and understanding.} This table compares OneEmo against three categories of models: open-source specialist models, open-source generalist models, and commercial generalist models. $\uparrow$ indicates that higher values are better for this metric. The best performance is shown in \textbf{bold}, and the second-best is \underline{underlined}. OneEmo outperforms other open-source generalist models and specialist models on most tasks. Despite having a smaller model size than closed-source models, OneEmo still achieves competitive results.}
\label{tab:main_6tasks}
\begin{tabular}{lcccccc|ccccc|c|c|c|ccc|c}
%\toprule[1.2pt]
\toprule
\multirow{4}{*}{\textbf{Models}} & \multirow{4}{*}{\textbf{Size}} & \multicolumn{5}{c}{\textbf{MSA}} & \multicolumn{5}{c}{\textbf{B-MER}} &  \multicolumn{1}{c}{\textbf{OV-MER}} &  \multicolumn{1}{c}{\textbf{MHU}} & \multicolumn{1}{c}{\textbf{MSU}} & \multicolumn{3}{c|}{\textbf{MIR}} & \textbf{Avg.}\\
\cmidrule(lr){3-18}
 &  & MOSI & MOSEI & SIMS & SIMS-v2 & \multirow{2}{*}{Avg.} & MER'23 & MER'24 & MELD & IEMOCAP & \multirow{2}{*}{Avg.} & OVMERD & UR-FUNNY & MUStARD & MIRec & MIRec2 &\multirow{2}{*}{Avg.} & \\
\cmidrule(lr){3-6} \cmidrule(lr){8-11} \cmidrule(lr){13-13} \cmidrule(lr){14-14} \cmidrule(lr){15-15} \cmidrule(lr){16-17} 
 &  & \multicolumn{4}{c}{WAF $\uparrow$} &  & \multicolumn{4}{c}{Hit Rate $\uparrow$} &  & EW $\uparrow$ & WAF $\uparrow$ & WAF $\uparrow$ & \multicolumn{2}{c}{WAF $\uparrow$} & & \\
\midrule
\multicolumn{19}{c}{\textit{Open-source Specialist Models}} \\
\midrule
R1-Omni~\cite{zhao2025r1} & 2.1B & 55.56 & 48.62 & 74.71 & 76.67 & 63.89 & 58.30 & 69.41 & 40.87 & 50.18 & 54.69 & 51.84 & - & - & - & - & - & -\\
Nano-EmoX~\cite{huang2026nano-emox} & 2.2B & 76.82 & 79.81 & 86.25 & 84.76 & \underline{81.91} & 79.09 & 77.94 & 56.55 & 60.12 & \textbf{68.43} & \textbf{64.75} & - & - &  58.17 & 47.27 &52.72 & -\\
AffectGPT~\cite{lian2025affectgpt} & 8.3B & 81.30 & 80.90 & 88.49 & 86.18 & \textbf{84.22} & 78.54 & 78.80 & 55.65 & 60.54 & \underline{68.38} & \underline{60.41} & - & - & - & - & -  & -\\

\midrule
\multicolumn{19}{c}{\textit{Open-source Generalist Models}} \\
\midrule
VidEmo-3B~\cite{zhang2025VidEmo} & 3.8B & 66.37 & 46.89 & 76.67 & 69.28 & 64.80 & 40.05 & 50.37 & 33.75 & 29.96 & 38.53 & 55.88 & 48.97 & 48.93 & 15.33 & 5.94 &10.64 & 44.63\\
VidEmo-7B~\cite{zhang2025VidEmo}  & 7.7B & 65.19 & 59.46 & 78.32 & 76.44 & 69.85 & 46.87 & 55.70 & 32.95 & 32.81 & 42.08 & 57.98 & 45.99 & 53.37 & 11.57 & 10.36 &10.97 & 46.71\\
MiniCPM-V-2.6~\cite{yao2024minicpm} & 8.0B & 74.96 & 57.44 & 74.85 & 75.04 & 70.57 & 46.67 & 45.31 & 40.27 & 36.31 & 42.14 & 50.04 & 61.71 & 46.63 & 32.78 & 23.98 & 28.38 & 49.91\\
InternVL3.5-4B~\cite{wang2025internvl35advancingopensourcemultimodal} & 4.7B  & 75.56 & 72.72 & 75.05 & 77.07 & 75.10 & 64.23 & 61.38 & 52.73 & 54.00  & 58.09 & 56.64 & 46.00 & 35.11 & 49.13 & 29.87 & 39.50 & 51.74 \\
Qwen3.5-4B~\cite{qwen2026qwen35} & 4.5B & 74.68 & 63.25 & 78.54 & 79.61 & 74.02 & 51.97  & 55.68 & 32.44 & 37.51 & 44.40 & 61.52 & 58.43 & 55.92 & 44.94 & 26.11 & 35.53 & 54.97\\
Qwen3.5-9B~\cite{qwen2026qwen35} & 9.7B & 68.60 & 64.42 & 80.17 & 77.97 & 72.79 & 52.64 & 64.51 & 41.26 & 43.97 & 50.60 & 60.10 & 54.25 & 53.28 & 47.64 & 33.60 & 40.62 & 55.27 \\
Qwen2.5-VL-7B~\cite{bai2025qwen25vl} & 7.7B & 74.10 & 58.35 & 78.65 & 77.43  & 72.13 & 59.81 & 69.14 & 48.05 & 50.53 & 56.89 & 55.61 & 53.28 & 54.66 & 53.28 & 33.54 & 43.41 & 56.00\\
Qwen3-VL-4B~\cite{bai2025qwen3vltechnicalreport} & 4.5B & 74.02 & 61.77 & 79.60 & 78.20 & 73.40 & 52.54 & 65.36 & 41.59 & 47.93 & 51.86 & 61.51 & \textbf{71.48} & 55.90 & 48.05 & 29.19 & 38.62 & 58.80\\
Cosmos3-Nano~\cite{nvidia2026cosmos3omnimodalworld}  & 16.0B  & 73.17 & 76.88 & 78.24 & 79.03 & 76.83 & 68.71 & 67.34 & 50.62 & 51.72 & 59.60  & 58.70 & 57.28 & 66.24 & 58.63 & 36.99 & 47.81 & 61.08\\
\textbf{OneEmo(+Curriculum)} & \textbf{4.5B} & 76.21 & 76.71 & 83.23 & 82.23 & \underline{79.60} & 70.51 & 73.12 & 53.05 & 58.69 & \underline{63.84} & \underline{65.40} & 68.80 & \underline{69.22} & 70.48 & 50.86 & \underline{60.67} & \underline{67.92}\\
\textbf{OneEmo(+Emo-Chord)} & \textbf{4.5B} & 75.88 & 78.13 & 85.54 & 84.98 & \textbf{81.13} & 76.03 & 80.21 & 55.06 & 63.15 & \textbf{68.61} & \textbf{68.54} & \underline{70.42} & \textbf{74.16} & 72.48 & 56.46 & \textbf{64.47} & \textbf{71.22}\\
\midrule

\multicolumn{19}{c}{\textit{Commercial Generalist Models}} \\
\midrule
GPT-5-mini~\cite{openai2025gpt5} & - & 80.64  & 63.17 & 77.25 & 77.88 & 74.74 & 60.49 & 58.43 & 53.36 & 46.04 & 54.58 & 58.80 & 78.47 & 62.62 & 65.21 & 43.21 & \underline{54.21} &63.90\\
Gemini-3.1-Pro~\cite{deepmind2026gemini31pro} & - & 76.59 & 54.98 & 84.07 & 84.89 & \underline{75.13} & 74.64 & 78.17 & 62.72 & 70.85 & \textbf{71.60} & \textbf{72.90} & \underline{85.46} & \underline{70.61} & 68.31 & 54.47 &\textbf{61.39} & \textbf{72.85}\\
MiMo-v2.5~\cite{xiaomi2026mimov25pro} & 310B & 83.75 & 66.11 & 88.05 & 87.58 & \textbf{81.37} & 73.21 & 81.32 & 52.24 & 61.77 & \underline{67.13} & \underline{65.22} & \textbf{87.04} & \textbf{74.70} & 62.20 & 43.96 &53.08 & \underline{71.42}\\

% \bottomrule[1.2pt]
\bottomrule
\end{tabular}%
}
\end{table*}

\section{Experimental Setup and Dataset}

\subsection{Implementation Details}

% We implement Emo-Chord utilizing the SFT and RL subsets of EmoWorld-130K, training on three RTX 4090 and two H800 GPUs. All phases employ the AdamW optimizer, a micro-batch size of 1, and 8 gradient accumulation steps. The off-policy cold-start phase spans 5 epochs with a learning rate of $1\times10^{-5}$. This is followed by a 2-epoch collaborative exploration-imitation phase, where the auxiliary loss balancing coefficient $\mu$ is progressively annealed from 0.5 to 0.02. For the GRPO, we set the learning rate to $2\times10^{-6}$, sampling group size to 8, and KL divergence coefficient to 0.06. Qwen2.5-7B~\cite{qwen2025qwen25technicalreport} serves as the reward model, with scaling factors for the format, thought, and answer rewards established at 0.5, 0.3, and 1.0, respectively.

OneEmo is trained on EmoWorld-130K using the Emo-Chord strategy using two NVIDIA H800 GPUs. All training phases employ the AdamW optimizer with a micro-batch size of 1 and 8 gradient accumulation steps. The off-policy cold-start phase spans 5 epochs with a learning rate of $1\times10^{-5}$. This is followed by a 2-epoch RL phase, during which the auxiliary loss balancing coefficient $\mu$ is annealed from 0.5 to 0.02. For GRPO, we set the learning rate to $2\times10^{-6}$, the sampling group size to 8, and the KL divergence coefficient to 0.06. The reward coefficients are set as follows: $\gamma_{f}=0.5, \gamma_{t}=0.3, \gamma_{a}=1.0$. In our main experiments, we adopt Qwen3.5-4B~\cite{qwen2026qwen35} as the backbone, unfreezing the multimodal adapter and fine-tuning the language backbone via LoRA. Additional results utilizing alternative backbones are presented in Section~\ref{sec:experiment_backbone}.

% \subsection{Benchmarks and Metrics}
% Table~\ref{tab:benchmarks} presents the benchmarks that we used for the assessment. For perception and understanding, we report Weighted Average F1 (WAF) for MSA and OV-MER, Hit Rate for B-MER, and Weighted F1 (WF1) for MIR, MSU, and MHU. For open-ended interactions, we adopt a 1-5 Likert-scale framework~\cite{2022CEM, zhao-etal-2024-esc} evaluated by three LLMs (GPT-4.1-mini~\cite{openai2025gpt41}, MiMo-v2.5-pro~\cite{xiaomi2026mimov25pro}, DeepSeek-v4-flash~\cite{deepseek2026v4}). Specifically, ERG is assessed on Empathy (Emp.), Contextual Coherence (Coh.), and Informativeness (Inf.), whereas ESC is measured by Emp., Skill use (Skill), and Overall effect (Ove.) dimension. Scores are Arithmetic mean-aggregated, alongside Randolph's $\kappa$ for inter-rater reliability. Additionally, blind pairwise human evaluations on 100 randomly selected instances by 3 psychology undergraduates validate Emp. and humanoid (Hum.) for ERG, alongside Ove. and Hum. for ESC.

\subsection{Benchmarks}

Table~\ref{tab:benchmarks} summarizes the evaluation benchmarks. For emotion perception, we adopt a cross-corpus evaluation protocol, ensuring that training and testing data are sourced from disjoint datasets. For emotion understanding and interaction, we adhere to the official data splits. The training and validation sets constitute EmoWorld-130K, while the test set is held out for final performance evaluation.

\subsection{Evaluation Metrics}
We adopt the official evaluation metrics for each task to ensure fair comparisons. For perception and understanding, we report the Weighted Average F1 (WAF) for MSA~\cite{zadeh2016mosi}, Hit Rate for B-MER~\cite{lian2025affectgpt}, Emotion Wheel-based metric~(EW) for OV-MER~\cite{lian2025ov}, and WAF for MIR, MSU, and MHU~\cite{MIntRec}. For interactions, we employ both automatic and human evaluations. 1) Automatic metrics utilize a 1–5 Likert scale~\cite{2022CEM, zhao-etal-2024-esc}, scored by diverse LLMs (GPT-4.1-mini, MiMo-v2.5-pro, DeepSeek-v4-flash) to mitigate potential LLM-as-a-judge bias. Specifically, ERG is evaluated on Empathy (Emp.), Contextual Coherence (Coh.), and Informativeness (Inf.), while ESC is assessed via Empathy (Emp.), Skill Use (Skill), and Overall Effect (Ove.). Scores are aggregated using arithmetic means, accompanied by Randolph’s $\kappa$ for inter-rater reliability. 2) For human evaluation, we conduct blind pairwise comparisons on 100 randomly sampled instances, rated by three psychology undergraduates. This validates Empathy (Emp.) and Humanoid Alignment (Hum.) for ERG, and Overall Effect (Ove.) and Humanoid Alignment (Hum.) for ESC.

\section{Results and Discussion}

\subsection{Main Results}
\textbf{Emotion Perception.} Table~\ref{tab:main_6tasks} presents the zero-shot evaluation results for MSA, B-MER, and OV-MER. OneEmo surpasses all open-source generalists, including 16B-scale multimodal generalists, \eg Cosmos3-Nano. In comparisons with commercial models, it outperforms models such as GPT-5-Mini and MiMo-v2.5 on B-MER and OV-MER, trailing by only 0.24 points on MSA despite a 70× parameter gap. Furthermore, OneEmo optimized via curriculum SFT exhibits highly competitive capabilities in mapping multimodal representational cues into emotional spaces.

\textbf{Emotion Understanding.} As shown in Table~\ref{tab:main_6tasks}, OneEmo demonstrates exceptional proficiency in conversational expression and social intent recognition. Notably, on complex 20- and 30-class intent recognition, it outperforms Gemini-3.1-Pro by a notable margin of 3.08 points. Operating at an efficient 4.5B parameter scale, OneEmo comprehensively eclipses both open-source affective specialists and generalist models across perception and understanding levels. Moreover, it remains highly competitive against top-tier commercial models, trailing Gemini-3.1-Pro by a mere 1.63 points. Unlike existing affective specialists, OneEmo goes beyond mere emotion perception; it seamlessly integrates multimodal cue extraction with the context-aware comprehension of human expressions and social intents. These robust outcomes are fundamentally attributed to the Emo-Chord training paradigm. By synergizing environmental exploration with expert imitation, this strategy empowers compact models to overcome parameter constraints, facilitating robust emotional reasoning capabilities.

\textbf{Emotion Interaction.} Table~\ref{tab:main_2tasks} outlines the LLM evaluation results for ERG and ESC across three dimensions. OneEmo outperforms all open-source baselines, achieving superior empathy in empathetic dialogues and a higher overall effect in emotional support scenarios. Following RL optimization, we observe a marginal decline in informativeness and contextual coherence. We attribute this variance to the human-annotated reference responses in EmoWorld-130K, which intrinsically lack the exhaustive, multi-solution enumerative formats commonly generated by commercial chatbots. However, showing detailed suggestions and proficient psychological techniques is not always appropriate in all situations, and it might even have a negative effect. Expectedly, top-tier commercial models retain an advantage in interaction-level tasks, benefiting from their massive parameter scales and extensive training corpora. Overall, OneEmo delivers highly competitive emotional interaction capabilities within a lightweight parameter footprint.

\begin{table}[t]
\centering
\footnotesize
\caption{{Automatic evaluation results on emotion interaction.}}
\label{tab:main_2tasks}
\resizebox{\columnwidth}{!}{%
\begin{tabular}{lccc|ccc|c}
\toprule
\multicolumn{1}{c}{\multirow{3}{*}{\textbf{Models}}} & \multicolumn{3}{c|}{\textbf{ERG}} & \multicolumn{3}{c|}{\textbf{ESC}} & \multirow{3}{*}{$\kappa$} \\
\cmidrule(lr){2-4} \cmidrule(lr){5-7}
 & Emp. $\uparrow$ & Coh. $\uparrow$ & Inf. & Emp. $\uparrow$ & Skill & Ove. $\uparrow$ & \\
\midrule
\multicolumn{8}{c}{\textit{Commercial Generalist Models}} \\
\midrule
GPT-5-mini & 4.40 & 4.94 & 4.61 & 4.27 & 4.71 & 4.33 & 0.45 \\
Gemini-3.1-Pro & 4.48 & 4.97 & 3.80 & 4.69 & 4.58 & 4.65 & 0.69 \\
MiMo-v2.5 & 4.56 & 4.86 & 3.78 & 4.56 & 4.50 & 4.52 & 0.57 \\
\midrule
\multicolumn{8}{c}{\textit{Open-source Generalist Models}} \\
\midrule
Qwen3.5-9B & 3.86 & 4.01 & 3.16 & \textbf{4.62} & \textbf{4.58} & \underline{4.51} & 0.45 \\
Qwen3-VL-4B & \underline{4.11} & 4.07 & 3.07 & 4.06 & 3.11 & 3.30 & 0.51 \\
Qwen3.5-4B & 3.96 & 4.22 & 3.00 & 4.40 & 4.21 & 4.20 & 0.46 \\
\textbf{OneEmo(+Curriculum)} &3.10 & \textbf{4.38} &\underline{3.21} &4.34 &4.07 &4.24 &0.50 \\
\textbf{OneEmo(+Emo-Chord)} & \textbf{4.25} & 4.16 &3.10 & \underline{4.51} & \underline{4.42} &\textbf{4.67} & 0.54\\
\bottomrule	
\end{tabular}%
}
\end{table}

\begin{table}[t]
\centering
\footnotesize
\caption{{Human evaluation results on emotion interaction.}}
\label{tab:main_human_eval}
\resizebox{\columnwidth}{!}{%
\begin{tabular}{lcrrr|crrr}
\toprule
\multirow{2}{*}{\textbf{OneEmo vs.}} & \multirow{2}{*}{\textbf{Dim}} & \multicolumn{3}{c|}{\textbf{ERG}} & \multirow{2}{*}{\textbf{Dim}}& \multicolumn{3}{c}{\textbf{ESC}} \\
\cmidrule(lr){3-5} \cmidrule(lr){7-9}
 &  & Win & Tie & Lose &  & Win & Tie & Lose \\
\midrule
\multirow{2}{*}{GPT-5-mini} & Emp. & 32.00 & 19.00  & 49.00 & Ove. & \textbf{57.00} & 12.00 & 31.00 \\
 & Hum. & \textbf{72.00} & 17.33 & 10.67 & Hum. & \textbf{60.67} & 18.67 & 20.67 \\
\midrule
\multirow{2}{*}{MiMo-v2.5} & Emp. & 41.67 & 7.33 & 51.00 & Ove. & 32.67 & \textbf{34.67} & 32.67 \\
 & Hum. & \textbf{70.33} & 10.67 & 19.00 & Hum. & 29.00 & \textbf{38.00} & 33.00 \\
 \midrule
\multirow{2}{*}{Qwen3.5-9B} & Emp. & \textbf{50.67} & 3.67 & 45.67 & Ove. & \textbf{71.33} & 3.67 & 25.00 \\
 & Hum. & \textbf{80.33} & 4.67 & 15.00 & Hum. & \textbf{84.33} & 6.67 & 9.00 \\
 \midrule
\multirow{2}{*}{Qwen3-VL-4B} & Emp. & \textbf{49.67} & 4.00 & 46.33 & Ove. & \textbf{45.33} & 37.67 & 17.00 \\
 & Hum. & \textbf{78.33} & 7.00 & 14.67 & Hum. & \textbf{50.33} & 39.00 & 10.67 \\
\bottomrule
\end{tabular}%
}
\begin{tablenotes}
\scriptsize
% \item We used a two-sided exact sign test, excluding ties, to test whether win and loss counts significantly deviated from a 50/50 distribution. 
\item The differences in win/loss counts were statistically significant across all dimensions ($p$-value \textless 0.01), except for a statistical tie against MiMo-v2.5.
\end{tablenotes}
\end{table}
%------------------------------------------------------------------------------------

Table~\ref{tab:main_human_eval} presents the results of the blind human evaluation. Across both ERG and ESC tasks, human annotators consistently highlight OneEmo's exceptional capability to deliver highly anthropomorphic interactions, achieving a remarkable peak win rate of 84.33\% in human-likeness. Nevertheless, OneEmo trails GPT-5-mini and MiMo-v2.5 on ERG task, it still exhibits highly competitive performance against massive proprietary models. In the complex ESC task, it decisively surpasses GPT-5-mini and achieves a statistical tie with the 310B-parameter MiMo-v2.5. Furthermore, when evaluated against state-of-the-art open-source baselines, our model establishes comprehensive dominance in both interaction scenarios. These subjective assessments compellingly validate that our proposed framework translates robustly into superior human-centric emotional engagements in the real-world.

%-------------------------------ablation tab 1--------------------------------------
% \begin{table}[!t]
% \caption{The experimental results of the reward function ablation study for the Emo-Chord framework.}
% \label{tab:reward_ablation}
% \centering
% \resizebox{\columnwidth}{!}{%
% \begin{tabular}{ccc|cccccccc}
% \toprule[1.2pt]
% $R_{\text{format}}$ & $R_{\text{thought}}$ & $R_{\text{answer}}$ & MSA & B-MER & OV-MER & MIR & MSU & MHU &ERG &ESC\\
% \midrule
%  &  & & 74.02 & 44.40 & 61.52 & 35.53 & 58.43 & 55.92 &3.73  & 4.27  \\
%            &            & \checkmark & 80.25 & 65.60 &63.64 & 57.40 & 68.02 &73.77 &   &    \\
% \checkmark & \checkmark &            & 80.36 & 63.68 & 66.05 & 59.68 & 61.57 &  72.87 &  &    \\
% \checkmark &            & \checkmark & 80.79 & 68.18 & 66.80 & 64.32 & 70.02 & 68.41 &  &      \\
%            & \checkmark & \checkmark & 80.81 & 65.16 & 65.46& 59.52 & 65.81 & 70.44 &  &     \\
% \checkmark & \checkmark & \checkmark &  81.13   & 68.61 & 68.54 & 64.47 & 70.42 & 74.16 &3.84  &4.53    \\
% \bottomrule[1.2pt]
% \end{tabular}
% }
% \end{table}

\begin{table}[!t]
\centering
\scriptsize
\setlength{\tabcolsep}{2.0pt}
\label{tab:reward_ablation}
\caption{Ablation study of reward functions.}
% \resizebox{\columnwidth}{!}{%
\begin{tabular}{ccc|cccccccc}
\toprule
$R_{\text{format}}$ & $R_{\text{thought}}$ & $R_{\text{answer}}$ & MSA & B-MER & OV-MER & MIR & MSU & MHU &ERG &ESC\\
\midrule
$\boldsymbol{\times}$ & $\boldsymbol{\times}$ & $\boldsymbol{\times}$ & 74.02 & 44.40 & 61.52 & 35.53 & 58.43 & 55.92 &3.73  & 4.27  \\
$\boldsymbol{\times}$ & $\boldsymbol{\times}$ & \checkmark & 80.25 & 65.60 &63.64 & 57.40 & 68.02 &73.77 & 3.75  & 4.03   \\
\checkmark & \checkmark & $\boldsymbol{\times}$ & 80.36 & 63.68 & 66.05 & 59.68 & 61.57 &  72.87 & 3.80 & 3.97    \\
\checkmark & $\boldsymbol{\times}$ & \checkmark & 80.79 & 68.18 & 66.80 & 64.32 & 70.02 & 68.41 & 3.62 & 4.25     \\
 $\boldsymbol{\times}$ & \checkmark & \checkmark & 80.81 & 65.16 & 65.46& 59.52 & 65.81 & 70.44 & 3.76 & 4.12    \\
\checkmark & \checkmark & \checkmark &  \textbf{81.13}   & \textbf{68.61} & \textbf{68.54} & \textbf{64.47} & \textbf{70.42} & \textbf{74.16} &\textbf{3.84}  &\textbf{4.53}    \\
\bottomrule
\end{tabular}

% }
\end{table}
%-------------------------------abltion tab 1--------------------------------------

\begin{table*}[h]
\scriptsize
\centering
\caption{Different model variants and their training details.}
\label{tab:training_strategies}
\begin{tabular}{p{1cm}p{6.6cm}p{8.6cm}}
\toprule
\textbf{Variants} & \textbf{Training Pipeline} & \textbf{Training Details} \\
\midrule
Vanilla & --- & Raw backbone \\
\midrule
SFT-v1 & Vanilla+SFT(P) & SFT on perception tasks \\
SFT-v2 & Vanilla+SFT(P+U) & SFT on perception and understanding tasks \\
SFT-v3 & Vanilla+SFT(P+U+I) & SFT on perception, understanding, and interaction tasks \\
SFT-v4 & Vanilla+CoT-free SFT(P+U+I) & SFT-v3 without explicit thinking supervision \\
SFT-v5 & Vanilla+Curriculum SFT(P+U+I) & SFT-v3 with two-stage curriculum learning \\
\midrule
RL-v1 & Vanilla+Curriculum SFT(S1)+RL & Initialized from the first stage of Curriculum SFT and optimized via GRPO \\
RL-v2 & Vanilla+Curriculum SFT(S2)+RL & Initialized from the second stage of Curriculum SFT and optimized via GRPO \\
RL-v3 & Vanilla+RL mix SFT & Direct GRPO training with auxiliary SFT loss \\
RL-v4 & Vanilla+Curriculum SFT(S1)+RL mix SFT \textbf{(Emo-Chord)} & Initialized from the first stage of Curriculum SFT and optimized via RL mix SFT \\
RL-v5 & Vanilla+Curriculum SFT(S2)+RL mix SFT & Initialized from the second stage of Curriculum SFT and optimized via RL mix SFT \\
\bottomrule
\end{tabular}
\end{table*}

\subsection{Impact of Reward Functions}

As shown in Table~\ref{tab:reward_ablation}, the full reward configuration yields comprehensive performance enhancements. Specifically, ablating the answer reward leads to a substantial average decline of 2.97 points across eight tasks. Similarly, removing the thought and format rewards results in average decrements of 1.16 and 2.58 points, respectively. While answer and format rewards directly guarantee outcome accuracy and structural compliance, the thought reward uniquely anchors multi-level capabilities. Specifically, it enforces multimodal factual consistency and reasoning coherence in perception tasks, while supervising user state modeling and strategy selection in interactive scenarios. By imposing these fine-grained semantic constraints, our design prevents reliance on spurious correlations. Consequently, final predictions are firmly grounded in logically sound reasoning processes, ultimately driving OneEmo's exceptional multi-task generalization.

%%%%%%%%%%%%%%%%%%%%%%%%%%%%%%%%%%%%%%fig5%%%%%%%%%%%%%%%%%%%%%%%%%%%%%%%%%%%%%%%%%%%%%%%%%%%%%%
\begin{figure}[t]
\centering
\includegraphics[width=\columnwidth]{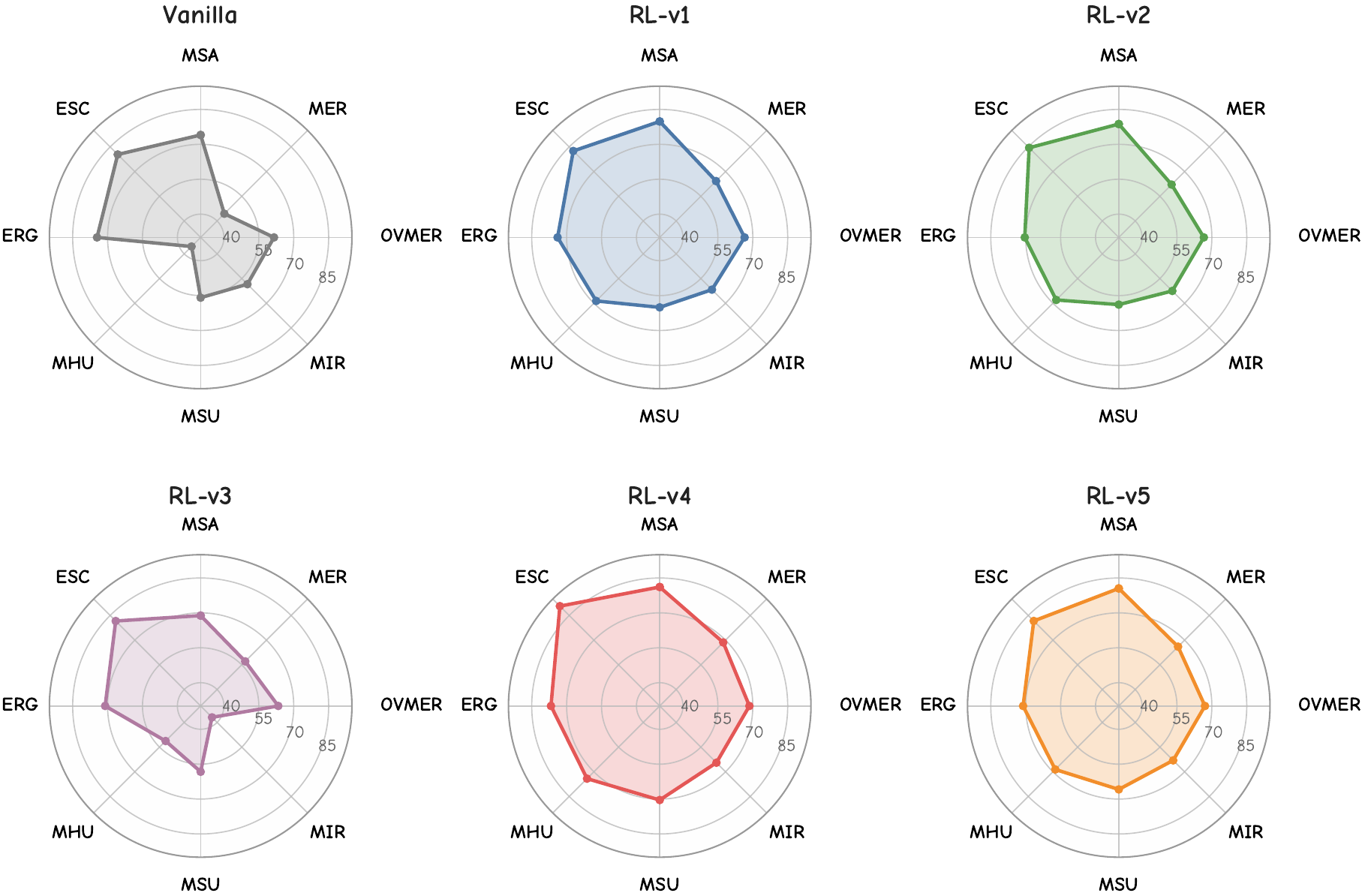}
\caption{
%Training Strategy Comparison. To facilitate a more intuitive comparison, we have linearly scaled the performance indicators of the ERG and ESC tasks by a factor of 20.
\textbf{Training strategy comparison.} We linearly scaled the evaluation metrics for ERG and ESC by a factor of 20 to align their numerical ranges with those of other tasks.}
\label{fig:ablation-training-strategy}
\end{figure}
%%%%%%%%%%%%%%%%%%%%%%%%%%%%%%%%%%%%%%%fig5%%%%%%%%%%%%%%%%%%%%%%%%%%%%%%%%%%%%%%%%%%%%%%%%%%%%%
\subsection{Impact of Different Training Strategy}
\label{sec:experiment_training_strategy}

We conduct comprehensive studies to validate the design choices of our training strategy. Table~\ref{tab:training_strategies} summarizes the training details and variants. Fig.~\ref{fig:ablation-training-strategy} presents the performance of diverse RL paradigms compared to the vanilla baseline. Directly applying joint RL and SFT optimization without a prior off-policy cold-start (RL-v3) yields severely degraded performance, particularly struggling on complex tasks like MIR and MHU. Without foundational multimodal grounding, the model fails to secure consistent reward signals across heterogeneous tasks, leading to unstable optimization. Furthermore, regarding the optimal transition point for RL, we observe that initializing exploration from the first stage of Curriculum SFT (RL-v1, RL-v4) consistently outperforms initialization from the fully completed second stage (RL-v2, RL-v5). We attribute this to policy entrenchment: exhaustive supervised curriculum training inadvertently over-solidifies sub-optimal behaviors, restricting the model’s exploratory capacity during subsequent RL phases. Among all paradigms, our proposed \textit{Emo-Chord} (RL-v4) achieves the optimal balance and highest aggregate gains. By coupling a strategically calibrated, limited cold-start with dynamic multi-task exploration synergized with expert data replay, \textit{Emo-Chord} effectively circumvents both reward sparsity and policy entrenchment, converging to vastly superior reasoning policies.

\begin{figure}[t]
\centering
\includegraphics[width=\columnwidth]{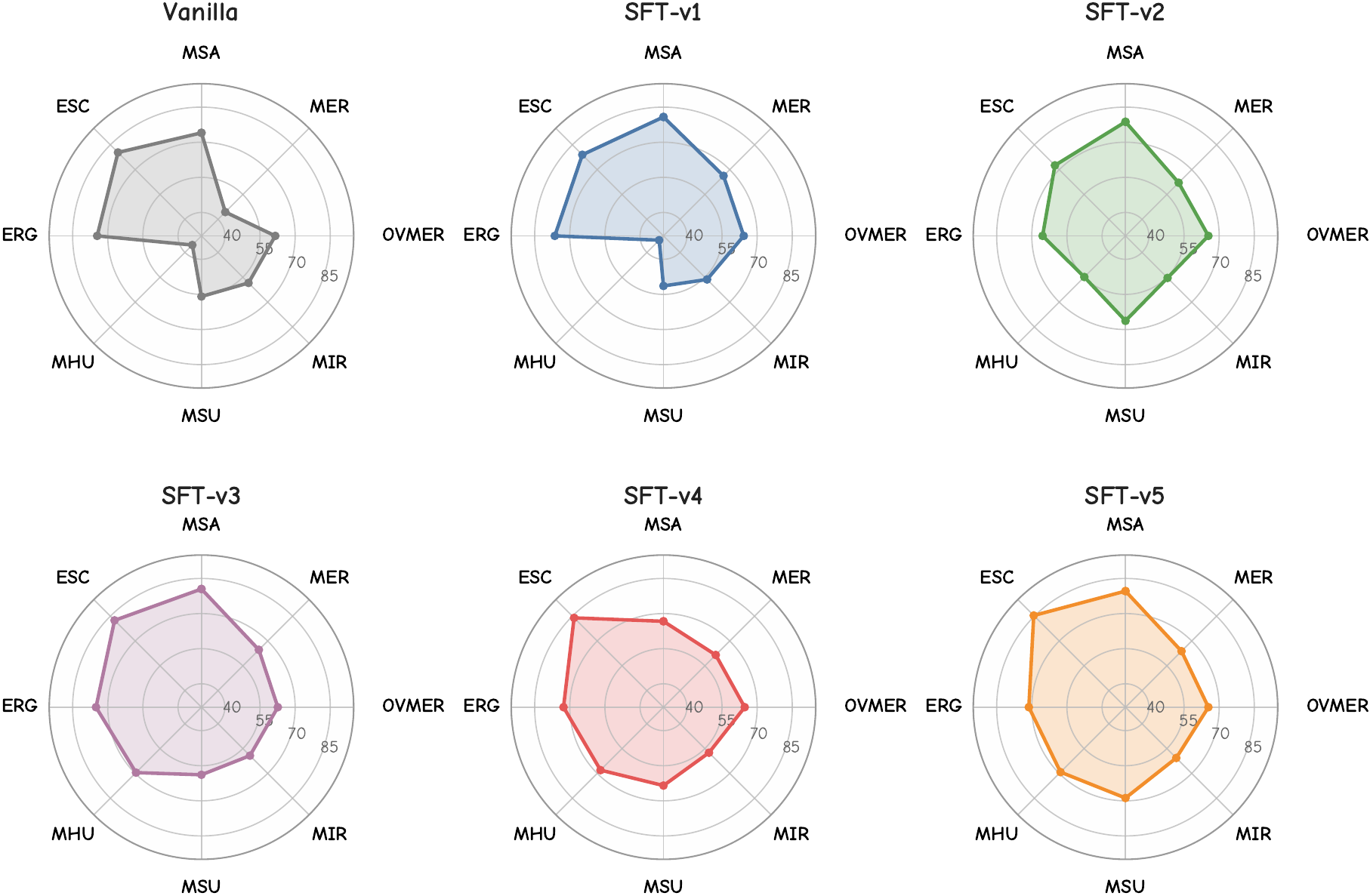}
\caption{\textbf{Task synergy analysis.} We conduct an ablation study to quantify the impact of varying training data compositions in SFT.}
\label{fig:ablation-task-synergy}
\end{figure}
%%%%%%%%%%%%%%%%%%%%%%%%%%%%%%%%%%%%fig6%%%%%%%%%%%%%%%%%%%%%%%%%%%%%%%%%%%%%%%%

\subsection{Task Synergy Analysis}
Fig~\ref{fig:ablation-task-synergy} illustrates the model performance under varying SFT data compositions. By examining the incremental integration of tasks from basic perception to empathetic generation (SFT-v1 to v3), we observe that naive joint training inherently suffers from task interference. While it enhances the actively trained tasks, it inadvertently degrades the performance of untrained domains, failing to foster effective cross-task synergy. Furthermore, the explicit thought-augmented variant (SFT-v3) maintains a consistent overall advantage over the direct-reasoning baseline (SFT-v4), suggesting the added value of incorporating explicit thought trajectories. Ultimately, by employing strategic data mixing and staged curriculum learning, SFT-v5 successfully navigates these bottlenecks. It demonstrates robust inter-task mutual promotion, effectively circumvents catastrophic forgetting, and yields a highly balanced comprehensive performance.

%%%%%%%%%%%%%%%%%%%%%%%%%%%%%%%%%%%%%%fig7%%%%%%%%%%%%%%%%%%%%%%%%%%%%%%%%%%%%%%%%%%%%%%%%%%%%%%
\begin{figure}[t]
\centering
\includegraphics[width=\columnwidth]{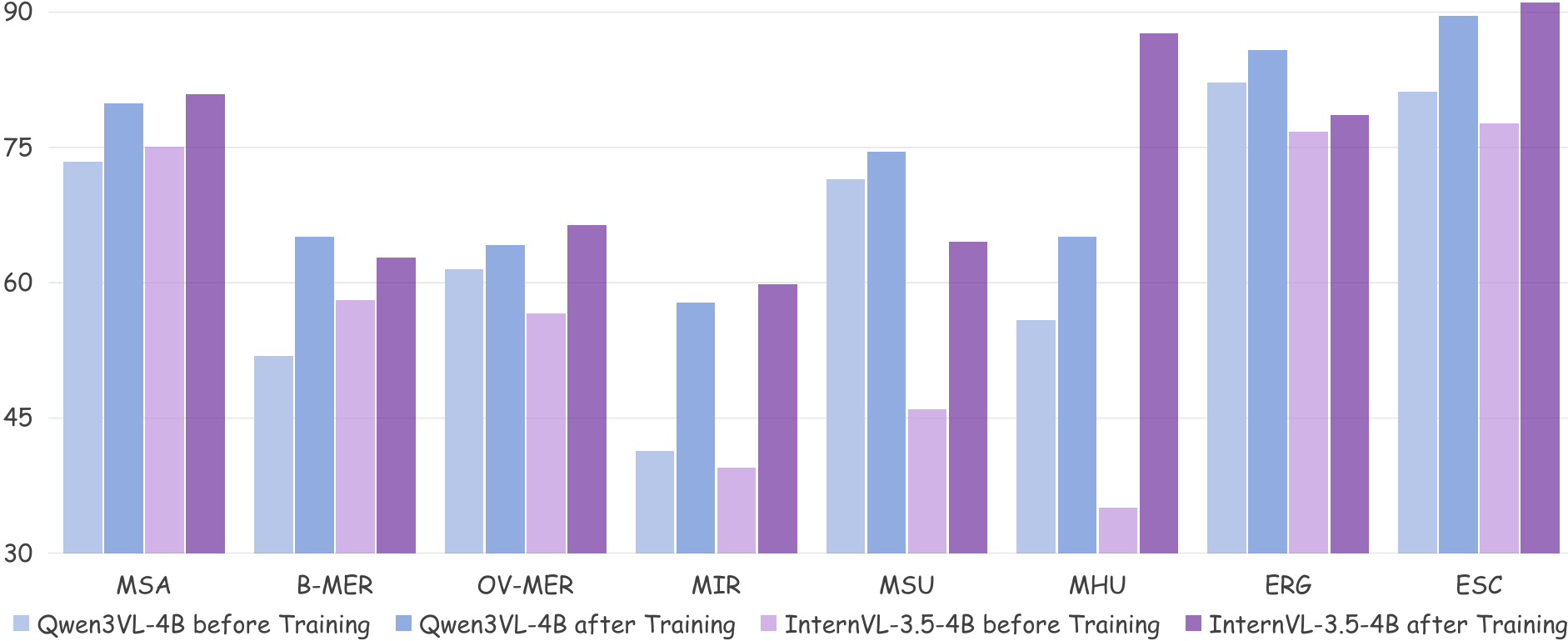}
\caption{
% Performance comparison of Qwen3-VL-4B and InternVL-3.5-4B before and after training on EmoWorld-130K. Post-training results reveal that both architectures achieve consistent performance enhancements across all eight affective tasks, validating the robust cross-model generalizability. 
\textbf{Performance across different backbones.} This figure compares the pre-training and post-training performance of two backbones: Qwen3-VL-4B and InternVL-3.5-4B. Experimental results demonstrate that our methods yields consistent improvements across all eight tasks, confirming the strong generalization capability of our approach.}
\label{fig:ablation-data-effect}
\end{figure}
%%%%%%%%%%%%%%%%%%%%%%%%%%%%%%%%%%%%%%%fig7%%%%%%%%%%%%%%%%%%%%%%%%%%%%%%%%%%%%%%%%%%%%%%%%%%%%%
\subsection{Generalization Across Different Backbones} 
\label{sec:experiment_backbone}

To evaluate the model-agnostic efficacy of our dataset, we apply LoRA fine-tuning to two distinct vision-language foundations: Qwen3-VL-4B and InternVL-3.5-4B. As illustrated by the pre- and post-training comparisons in Fig.~\ref{fig:ablation-data-effect}, both architectures exhibit strict Pareto improvements across the evaluated affective tasks. This compellingly demonstrates the robust cross-model generalizability of EmoWorld-130K. Regardless of the underlying architecture, our dataset systematically elevates the models' competencies from foundational perception to deep comprehension, ultimately unlocking highly reliable empathetic interaction capabilities while successfully avoiding inter-task capability conflicts.

\section{Conclusion}
% We present EmoWorld-130K, a comprehensive multimodal dataset that bridges the scarcity of explicit reasoning trajectories across multi-level affective tasks. Furthermore, we propose Emo-Chord, a unified multi-task RL strategy. By coupling a collaborative online-offline policy optimization process with synergistic reward allocation, Emo-Chord successfully mitigates task conflicts and fosters deep inter-task synergy. Consequently, our parameter-efficient OneEmo models deliver highly competitive performance across a broad spectrum of affective benchmarks. Together, these contributions substantially advance the interpretability and generalization of unified artificial emotional intelligence.

This paper introduces \textbf{OneEmo}, a unified reasoning model designed explicitly for emotional intelligence. To facilitate this, we present \textbf{EmoWorld-130K}, a comprehensive dataset featuring explicit reasoning trajectories across eight typical emotion tasks, and \textbf{Emo-Chord}, a novel multi-task reinforcement learning strategy that integrates online-offline collaborative policy optimization with well-designed reward allocation. Extensive experiments demonstrate that OneEmo consistently surpasses same-scale baselines on most benchmarks, validating the efficacy of explicit reasoning and inter-task synergy. Furthermore, comprehensive ablation studies verify the contribution of each module and elucidate the impact of distinct reward functions. This work marks an important step toward advancing unified artificial emotional intelligence.

\section{Ethics Statement}
EmoWorld-130K is derived exclusively from publicly licensed datasets without involving new human-subject data collection, and all human annotations were conducted by fairly compensated volunteers under strict anonymity and informed consent. Furthermore, although our emotional support tasks leverage clinical frameworks as reasoning scaffolding, OneEmo remains strictly a research prototype and must not substitute for professional psychiatric diagnosis or treatment.  We explicitly advise against its unsupervised clinical deployment. Consequently, any real-world application necessitates the integration of strict crisis-referral guardrails, and both the dataset and model will be released under a restrictive license rigorously enforcing these ethical and safety boundaries.

\section{Limitations and Future Works}
While our work represents a substantial stride toward unified artificial emotional intelligence, certain limitations remain. First, emotion is inherently context-sensitive and culturally diverse in its behavioral manifestations. Current explainable datasets often rely on scripted cinematic data; thus, future benchmarks must prioritize authentic, in-the-wild interactions. Second, although our framework effectively bridges perception, understanding, and interaction, real-world deployments demand an even broader task spectrum, such as continuous emotion forecasting and long-term empathetic companionship. Future research will focus on collecting diverse real-world data, fusing physiological and behavioral signals, and integrating these advanced capabilities to construct more robust, generalized affective agents.

\section*{Acknowledgments}
This work is supported by the Open Research Fund of the State Key Laboratory of Multimodal Artificial Intelligence Systems (MAIS2026003), the Innovative Team Project in Guangdong Province (2025KCXTD04), and Fujian Province Natural Science Foundation~(2026J001409).



\bibliographystyle{IEEEtran}
\bibliography{main}

@article{zhang2025emotional,
  title={Emotional Artificial Intelligence in Education: A Systematic Review and Meta-Analysis},
  author={Zhang, Heng and Liu, Yuhan and Jiang, Meilin and Chen, Juanjuan and Wang, Minhong and Paas, Fred},
  journal={Educational Psychology Review},
  volume={37},
  number={1},
  pages={106},
  year={2025},
  publisher={Springer},
}

@article{KumarM23,
  author       = {Himanshu Kumar and  A. Martin},
  title        = {Artificial Emotional Intelligence: Conventional and deep learning approach},
  journal      = {Expert Syst. Appl.},
  volume       = {212},
  pages        = {118651},
  year         = {2023}
}

@article{Seaborn2021,
  title={Voice in Human–Agent Interaction},
  author={Seaborn, Katie and Miyake, Norihisa and Pennefather, Peter and Otake-Matsuura, Mihoko},
  year={2021},
  journal={ACM Comput. Surv.},
  volume={54},
  number={4},
  pages={1-43}
}

@inproceedings{cheng2024emotion,
  title={Emotion-llama: Multimodal emotion recognition and reasoning with instruction tuning},
  author={Cheng, Zebang and Cheng, Zhi-Qi and He, Jun-Yan and Wang, Kai and Lin, Yuxiang and Lian, Zheng and Peng, Xiaojiang and Hauptmann, Alexander},
  booktitle={Advances in Neural Information Processing Systems 2024},
  volume={37},
  pages={110805--110853},
  year={2024}
}

@inproceedings{lian2025affectgpt,
  title={AffectGPT: A New Dataset, Model, and Benchmark for Emotion Understanding with Multimodal Large Language Models},
  author={Lian, Zheng and Chen, Haoyu and Chen, Lan and Sun, Haiyang and Sun, Licai and Ren, Yong and Cheng, Zebang and Liu, Bin and Liu, Rui and Peng, Xiaojiang and others},
  booktitle={Proceedings of the International Conference on Machine Learning (ICML) (Oral, Top 1\%)},
  year={2025}
}

@inproceedings{10.1145/3746027.3762029,
author = {Lin, Ronghao and Shen, Shuai and Hu, Weipeng and He, Qiaolin and Xiong, Aolin and Huang, Li and Hu, Haifeng and Tan, Yap-peng},
title = {E3RG: Building Explicit Emotion-driven Empathetic Response Generation System with Multimodal Large Language Model},
year = {2025},
isbn = {9798400720352},
publisher = {Association for Computing Machinery},
booktitle = {Proceedings of the 33rd ACM International Conference on Multimedia},
pages = {14006–14013},
series = {MM '25}
}

@article{hu2025beyond,
  title={Beyond Empathy: Integrating Diagnostic and Therapeutic Reasoning with Large Language Models for Mental Health Counseling},
  author = {Hu, He and Zhou, Yucheng and Si, Juzheng and Wang, Qianning and Zhang, Hengheng and Ren, Fuji and Ma, Fei and Cui, Laizhong},
  journal={arXiv preprint arXiv:2505.15715},
  year={2025}
}

@article{deepseek-r1,
  title={DeepSeek-R1: Incentivizing Reasoning Capability in LLMs via Reinforcement Learning},
  author={DeepSeek-AI},
  year={2025},
  journal={arXiv preprint arXiv:2501.12948},
}

@article{zhao2025r1,
  title={R1-omni: Explainable omni-multimodal emotion recognition with reinforcement learning},
  author={Zhao, Jiaxing and Wei, Xihan and Bo, Liefeng},
  journal={arXiv preprint arXiv:2503.05379},
  year={2025}
}

@inproceedings{Rha_Yeo_Kim_Ro_2026, 
title={Emotion-Coherent Reasoning for Multimodal LLMs via Emotional Rationale Verifier}, 
author={Fang, Yiyang and Huang, Wenke and Fu, Pei and Yang, Yihao and Su, Kehua and Luo, Zhenbo and Luan, Jian and Ye, Mang},
volume={40}, 
booktitle={Proceedings of the AAAI Conference on Artificial Intelligence},
year={2026}, 
pages={2029–2037} 
}

@inproceedings{Fang2026,
  title={EMO-R3: Reflective Reinforcement Learning for Emotional Reasoning in Multimodal Large Language Models},
  author={Fang, Yiyang and Huang, Wenke and Fu, Pei and Yang, Yihao and Su, Kehua and Luo, Zhenbo and Luan, Jian and Ye, Mang},
  booktitle={2026 IEEE/CVF Conference on Computer Vision and Pattern Recognition (CVPR)},
  year={2026}
}

@inproceedings{Huy_Multimood,
title={Reinforce Trustworthiness in Multimodal Emotional Support System},
author={Huy M. Le and Dat Tien Nguyen and Ngan T. T. Vo and Tuan D. Q. Nguyen and Nguyen Binh Le and Duy Minh Ho Nguyen and Daniel Sonntag and Lizi Liao and Binh T. Nguyen},
volume={40},
booktitle={Proceedings of the AAAI Conference on Artificial Intelligence},
year={2026},
pages={31474–31482}
}

@inproceedings{zhang2025VidEmo,
  author = {Zhang, Zhicheng and Wang, Weicheng and Zhu, Yongjie and Qin, Wenyu and Wan, Pengfei and Zhang, Di and Yang, Jufeng},
  title = {VidEmo: Affective-Tree Reasoning for Emotion-Centric Video Foundation Models},
  booktitle = {Advances in Neural Information Processing Systems},
  year = {2025}
}

@InProceedings{huang2026nano-emox,
    author    = {Huang, Jiahao and Lin, Fengyan and Yang, Xuechao and Feng, Chen and Zhu, Kexin and Yang, Xu and Chen, Zhide},
    title     = {Nano-EmoX: Unifying Multimodal Emotional Intelligence from Perception to Empathy},
    booktitle = {Proceedings of the IEEE/CVF Conference on Computer Vision and Pattern Recognition (CVPR)},
    month     = {June},
    year      = {2026},
    pages     = {22986-22997}
}

@article{yang2025omni,
  title={Omni-emotion: Extending video mllm with detailed face and audio modeling for multimodal emotion analysis},
  author={Yang, Qize and Bai, Detao and Peng, Yi-Xing and Wei, Xihan},
  journal={arXiv preprint arXiv:2501.09502},
  year={2025}
}

@article{huang2025emotion,
  title={Emotion-Qwen: Training Hybrid Experts for Unified Emotion and General Vision-Language Understanding},
  author={Huang, Dawei and Li, Qing and Yan, Chuan and Cheng, Zebang and Huang, Yurong and Li, Xiang and Li, Bin and Wang, Xiaohui and Lian, Zheng and Peng, Xiaojiang},
  journal={arXiv preprint arXiv:2505.06685},
  year={2025}
}

@INPROCEEDINGS{zhou2026HIER,
  title={Evolutionary Multimodal Reasoning via Hierarchical Semantic Representation for Intent Recognition},
  author={Qianrui Zhou and Hua Xu and Yunjin Gu and Yifan Wang and Songze Li and Hanlei Zhang},
  booktitle={2026 IEEE/CVF Conference on Computer Vision and Pattern Recognition (CVPR)},  
  year={2026},
}

@inproceedings{10.1145/3701716.3718371,
author = {Gao, Tianhong and Shen, Genhang and Wu, Yuxuan and Feng, Zunlei and Zhang, Jinshan and Zhou, Sheng},
title = {EcomMIR: Towards Intelligent Multimodal Intent Recognition in E-Commerce Dialogue Systems},
year = {2025},
isbn = {9798400713316},
publisher = {Association for Computing Machinery},
address = {New York, NY, USA},
booktitle = {Companion Proceedings of the ACM on Web Conference 2025},
pages = {3049–3052},
series = {WWW '25}
}

@article{Lian2025affectgptr1,
  title={Affectgpt-r1: Leveraging reinforcement learning for open-vocabulary multimodal emotion recognition},
  author={Lian, Zheng and Zhang, Fan and Zhang, Yazhou and Tao, Jianhua and Liu, Rui and Chen, Haoyu and Li, Xiaobai and He, Bin},
  journal={arXiv preprint arXiv:2508.01318},
  year={2025}
}

@article{10899840,
  author={Ma, Hui and Zhang, Bo and Xu, Bo and Wang, Jian and Lin, Hongfei and Sun, Xiao},
  journal={IEEE Transactions on Affective Computing}, 
  title={Empathy Level Alignment via Reinforcement Learning for Empathetic Response Generation}, 
  year={2025},
  volume={16},
  number={3},
  pages={1873-1884},
  }

@article{2024DeepSeekMath,
  title={DeepSeekMath: Pushing the Limits of Mathematical Reasoning in Open Language Models},
  author={ Shao, Zhihong  and  Wang, Peiyi  and  Zhu, Qihao  and  Xu, Runxin  and  Song, Junxiao  and  Bi, Xiao  and  Zhang, Haowei  and  Zhang, Mingchuan  and  Li, Y. K.  and  Wu, Y. },
  journal={arXiv preprint arXiv:2402.03300},
  year={2024}
}

@article{facial_r12026, title={Facial-R1: Aligning Reasoning and Recognition for Facial Emotion Analysis}, 
volume={40}, 
number={32}, 
journal={Proceedings of the AAAI Conference on Artificial Intelligence}, 
author={Wu, Jiulong and Shen, Yucheng and Yan, Lingyong and Sun, Haixin and Xia, Deguo and Huang, Jizhou and Cao, Min}, 
year={2026}, 
month={Mar.}, 
pages={26939–26947} 
}

@ARTICLE{11347465,
  author={Lian, Zheng and Sun, Licai and Ren, Yong and Gu, Hao and Sun, Haiyang and Chen, Lan and Liu, Bin and Tao, Jianhua},
  journal={IEEE Transactions on Pattern Analysis and Machine Intelligence}, 
  title={MERBench: A Unified Evaluation Benchmark for Multimodal Emotion Recognition}, 
  year={2026},
  volume={48},
  number={5},
  pages={5793-5810}
  }

@inproceedings{lian2023mer,
  title = {MER 2023: Multi-label Learning, Modality Robustness, and Semi-Supervised Learning},
  author = {Zheng Lian and Haiyang Sun and Licai Sun and Kang Chen and Mingyu Xu and Kexin Wang and Ke Xu and Yu He and Ying Li and Jinming Zhao and Ye Liu and Bin Liu and Jiangyan Yi and Meng Wang and Erik Cambria and Guoying Zhao and Björn W. Schuller and Jianhua Tao},
  booktitle={Proceedings of the 31st ACM international conference on multimedia},
  pages={9610--9614},
  year={2023}
}

@inproceedings{lian2024mer,
  title = {MER 2024: Semi-Supervised Learning, Noise Robustness, and Open-Vocabulary Multimodal Emotion Recognition},
  author = {Zheng Lian and Haiyang Sun and Licai Sun and Zhuofan Wen and Siyuan Zhang and Shun Chen and Hao Gu and Jinming Zhao and Ziyang Ma and Xie Chen and Jiangyan Yi and Rui Liu and Kele Xu and Bin Liu and Erik Cambria and Guoying Zhao and Björn W. Schuller and Jianhua Tao},
  booktitle={Proceedings of the 2nd International Workshop on Multimodal and Responsible Affective Computing},
  pages={41--48},
  year={2024}
}

@article{busso2008iemocap,
  title={IEMOCAP: Interactive emotional dyadic motion capture database},
  author={Busso, Carlos and Bulut, Murtaza and Lee, Chi-Chun and Kazemzadeh, Abe and Mower, Emily and Kim, Samuel and Chang, Jeannette N and Lee, Sungbok and Narayanan, Shrikanth S},
  journal={Language resources and evaluation},
  volume={42},
  number={4},
  pages={335--359},
  year={2008},
  publisher={Springer}
}

@article{zadeh2016mosi,
  author    = {Amir Zadeh and Rowan Zellers and Eli Pincus and Louis-Philippe Morency},
  title     = {MOSI: Multimodal Corpus of Sentiment Intensity and Subjectivity Analysis in Online Opinion Videos},
  journal   = {IEEE Intelligent Systems},
  year      = {2016},
  volume    = {31},
  number    = {6},
  pages     = {82--88},
}

@inproceedings{zadeh2018multimodal,
  title={Multimodal language analysis in the wild: Cmu-mosei dataset and interpretable dynamic fusion graph},
  author={Bagher Zadeh, AmirAli and Liang, Paul Pu and Poria, Soujanya and Cambria, Erik and Morency, Louis-Philippe},
  booktitle={Proceedings of the 56th Annual Meeting of the Association for Computational Linguistics (Volume 1: Long Papers)},
  pages={2236--2246},
  year={2018}
}

@inproceedings{yu2020ch,
  title={Ch-sims: A chinese multimodal sentiment analysis dataset with fine-grained annotation of modality},
  author={Yu, Wenmeng and Xu, Hua and Meng, Fanyang and Zhu, Yilin and Ma, Yixiao and Wu, Jiele and Zou, Jiyun and Yang, Kaicheng},
  booktitle={Proceedings of the 58th annual meeting of the association for computational linguistics},
  pages={3718--3727},
  year={2020}
}

@inproceedings{liu2022make,
  title={Make acoustic and visual cues matter: Ch-sims v2. 0 dataset and av-mixup consistent module},
  author={Liu, Yihe and Yuan, Ziqi and Mao, Huisheng and Liang, Zhiyun and Yang, Wanqiuyue and Qiu, Yuanzhe and Cheng, Tie and Li, Xiaoteng and Xu, Hua and Gao, Kai},
  booktitle={Proceedings of the 2022 international conference on multimodal interaction},
  pages={247--258},
  year={2022}
}

@inproceedings{zhang2025towards,
  title={Towards multimodal empathetic response generation: A rich text-speech-vision avatar-based benchmark},
  author={Zhang, Han and Meng, Zixiang and Luo, Meng and Han, Hong and Liao, Lizi and Cambria, Erik and Fei, Hao},
  booktitle={Proceedings of the ACM on Web Conference 2025},
  pages={2872--2881},
  year={2025}
}

@inproceedings{MIntRec,
   author = {Zhang, Hanlei and Xu, Hua and Wang, Xin and Zhou, Qianrui and Zhao, Shaojie and Teng, Jiayan},
   title = {MIntRec: A New Dataset for Multimodal Intent Recognition},
   year = {2022},
   booktitle = {Proceedings of the 30th ACM International Conference on Multimedia},
   pages = {1688–1697},
}

@inproceedings{zhang2024mintrec,
  title={MIntRec2.0: A Large-scale Benchmark Dataset for Multimodal Intent Recognition and Out-of-scope Detection in Conversations},
  author={Hanlei Zhang and Xin Wang and Hua Xu and Qianrui Zhou and Kai Gao and Jianhua Su and jinyue Zhao and Wenrui Li and Yanting Chen},
  booktitle={The Twelfth International Conference on Learning Representations},
  year={2024},
  url={https://openreview.net/forum?id=nY9nITZQjc}
}

@article{2022CEM,
  title={CEM: Commonsense-Aware Empathetic Response Generation},
  author={ Sabour, Sahand  and  Zheng, Chujie  and  Huang, Minlie },
  journal={Proceedings of the AAAI Conference on Artificial Intelligence},
  volume={36},
  number={10},
  pages={11229-11237},
  year={2022},
}

@inproceedings{zhao-etal-2024-esc,
    title = "{ESC}-Eval: Evaluating Emotion Support Conversations in Large Language Models",
    author = "Zhao, Haiquan  and
      Li, Lingyu  and
      Chen, Shisong  and
      Kong, Shuqi  and
      Wang, Jiaan  and
      Huang, Kexin  and
      Gu, Tianle  and
      Wang, Yixu  and
      Wang, Jian  and
      Dandan, Liang  and
      Li, Zhixu  and
      Teng, Yan  and
      Xiao, Yanghua  and
      Wang, Yingchun",
    editor = "Al-Onaizan, Yaser  and
      Bansal, Mohit  and
      Chen, Yun-Nung",
    booktitle = "Proceedings of the 2024 Conference on Empirical Methods in Natural Language Processing",
    month = nov,
    year = "2024",
    address = "Miami, Florida, USA",
    publisher = "Association for Computational Linguistics",
    pages = "15785--15810",
}

@inproceedings{hasan-etal-2019-ur,
    title = "{UR}-{FUNNY}: A Multimodal Language Dataset for Understanding Humor",
    author = "Hasan, Md Kamrul  and
      Rahman, Wasifur  and
      Bagher Zadeh, AmirAli  and
      Zhong, Jianyuan  and
      Tanveer, Md Iftekhar  and
      Morency, Louis-Philippe  and
      Hoque, Mohammed (Ehsan)",
    editor = "Inui, Kentaro  and
      Jiang, Jing  and
      Ng, Vincent  and
      Wan, Xiaojun",
    booktitle = "Proceedings of the 2019 Conference on Empirical Methods in Natural Language Processing and the 9th International Joint Conference on Natural Language Processing (EMNLP-IJCNLP)",
    month = nov,
    year = "2019",
    address = "Hong Kong, China",
    publisher = "Association for Computational Linguistics",
    pages = "2046--2056",
}

@misc{xiaomi2026mimov25pro,
  title={MiMo-V2.5-Pro: Model Card},
  author={{Xiaomi MiMo Team}},
  howpublished={\url{https://huggingface.co/XiaomiMiMo/MiMo-V2.5-Pro}},
  note={Accessed: 2026-06-06},
  year={2026},
}

@article{bai2025qwen25vl,
  title={Qwen2.5-VL Technical Report},
  author={Bai, Shuai and Chen, Keqin and Liu, Xuejing and Wang, Jialin and Ge, Wenbin and Song, Sibo and Dang, Kai and Wang, Peng and Wang, Shijie and Tang, Jun and Zhong, Humen and Zhu, Yuanzhi and Yang, Mingkun and Li, Zhaohai and Wan, Jianqiang and Wang, Pengfei and Ding, Wei and Fu, Zheren and Xu, Yiheng and Ye, Jiabo and Zhang, Xi and Xie, Tianbao and Cheng, Zesen and Zhang, Hang and Yang, Zhibo and Xu, Haiyang and Lin, Junyang},
  journal={arXiv preprint arXiv:2502.13923},
  year={2025}
}

@article{yao2024minicpm,
  title={MiniCPM-V: A GPT-4V Level MLLM on Your Phone},
  author={Yao, Yuan and Yu, Tianyu and Zhang, Ao and Wang, Chongyi and Cui, Junbo and Zhu, Hongji and Cai, Tianchi and Li, Haoyu and Zhao, Weilin and He, Zhihui and others},
  journal={arXiv preprint arXiv:2408.01800},
  year={2024}
}

@misc{openai2025gpt5,
  title={GPT-5 System Card},
  author={{OpenAI}},
  year={2025},
  howpublished={\url{https://cdn.openai.com/gpt-5-system-card.pdf}},
  note={Accessed: 2025-08-07}
}

@misc{deepmind2026gemini31pro,
  title={Gemini 3.1 Pro: A Smarter Model for Your Most Complex Tasks},
  author={{Google DeepMind}},
  year={2026},
  howpublished={\url{https://deepmind.google/models/gemini/pro}},
  note={Accessed: 2026-06-06}
}

@misc{qwen2026qwen35,
  title={Qwen3.5: Towards Native Multimodal Agents},
  author={{Qwen Team}},
  year={2026},
  howpublished={\url{https://qwen.ai/blog?id=qwen3.5}},
  note={Accessed: 2026-06-06}
}

@misc{bytedance2026seed2,
  title={Seed2.0: Towards Intelligence Frontier for Real-World Complex Tasks},
  author={{ByteDance Seed}},
  year={2026},
  howpublished={\url{https://lf3-static.bytednsdoc.com/obj/eden-cn/lapzild-tss/ljhwZthlaukjlkulzlp/seed2/0214/Seed2.0%20Model%20Card.pdf}},
  note={Model Card. Accessed: 2026-02-14}
}

@article{ekman1992argument,
  title={An argument for basic emotions},
  author={Ekman, Paul},
  journal={Cognition and Emotion},
  volume={6},
  number={3--4},
  pages={169--200},
  year={1992},
  publisher={Taylor \& Francis},
  doi={10.1080/02699939208411068}
}

@incollection{suls1972twostage,
  title={A two-stage model for the appreciation of jokes and cartoons: An information-processing analysis},
  author={Suls, Jerry M.},
  booktitle={The Psychology of Humor: Theoretical Perspectives and Empirical Issues},
  editor={Goldstein, Jeffrey H. and McGhee, Paul E.},
  pages={81--100},
  year={1972},
  publisher={Academic Press},
  address={New York}
}

@article{clark1984pretense,
  title={On the pretense theory of irony},
  author={Clark, Herbert H. and Gerrig, Richard J.},
  journal={Journal of Experimental Psychology: General},
  volume={113},
  number={1},
  pages={121--126},
  year={1984},
  publisher={American Psychological Association},
  doi={10.1037/0096-3445.113.1.121}
}

@book{lazarus1991emotion,
  title={Emotion and Adaptation},
  author={Lazarus, Richard S.},
  year={1991},
  publisher={Oxford University Press},
  address={New York}
}

@book{goldstein1972psychology,
  title={The Psychology of Humor: Theoretical Perspectives and Empirical Issues},
  editor={Goldstein, Jeffrey H. and McGhee, Paul E.},
  year={1972},
  publisher={Academic Press},
  address={New York}
}

@book{beck1976cognitive,
  title={Cognitive Therapy and the Emotional Disorders},
  author={Beck, Aaron T.},
  year={1976},
  publisher={International Universities Press},
  address={New York}
}

@book{hayes1999act,
  title={Acceptance and Commitment Therapy: An Experiential Approach to Behavior Change},
  author={Hayes, Steven C. and Strosahl, Kathleen D. and Wilson, Kelly G.},
  year={1999},
  publisher={Guilford Press},
  address={New York}
}

@book{apa2013dsm5,
  title={Diagnostic and Statistical Manual of Mental Disorders},
  author={{American Psychiatric Association}},
  edition={5th},
  year={2013},
  publisher={American Psychiatric Association Publishing},
  address={Washington, DC}
}

@book{who2022icd11,
  title={International Classification of Diseases for Mortality and Morbidity Statistics},
  author={{World Health Organization}},
  edition={11th Revision},
  year={2022},
  publisher={World Health Organization},
  address={Geneva, Switzerland}
}

@inproceedings{ijcai2024p0695,
  title     = {ECR-Chain: Advancing Generative Language Models to Better Emotion-Cause Reasoners through Reasoning Chains},
  author    = {Huang, Zhaopei and Zhao, Jinming and Jin, Qin},
  booktitle = {Proceedings of the Thirty-Third International Joint Conference on
               Artificial Intelligence, {IJCAI-24}},
  publisher = {International Joint Conferences on Artificial Intelligence Organization},
  editor    = {Kate Larson},
  pages     = {6288--6296},
  year      = {2024},
  month     = {8},
  note      = {Main Track},
}

@article{dai2025psycher1reliablepsychologicalllms,
      title={Psyche-R1: Towards Reliable Psychological LLMs through Unified Empathy, Expertise, and Reasoning}, 
      author={Chongyuan Dai and Jinpeng Hu and Hongchang Shi and Zhuo Li and Xun Yang and Meng Wang},
      year={2025},
      journal={arXiv preprint arXiv:2508.10848}, 
}

@inproceedings{jiang2020dfew,
  title={Dfew: A large-scale database for recognizing dynamic facial expressions in the wild},
  author={Jiang, Xingxun and Zong, Yuan and Zheng, Wenming and Tang, Chuangao and Xia, Wanchuang and Lu, Cheng and Liu, Jiateng},
  booktitle={Proceedings of the 28th ACM International Conference on Multimedia},
  pages={2881--2889},
  year={2020}
}

@inproceedings{lian2025mer,
  title={Mer 2025: When affective computing meets large language models},
  author={Lian, Zheng and Liu, Rui and Xu, Kele and Liu, Bin and Liu, Xuefei and Zhang, Yazhou and Liu, Xin and Li, Yong and Cheng, Zebang and Zuo, Haolin and others},
  booktitle={Proceedings of the 33th ACM International Conference on Multimedia},
  year={2025}
}

@inproceedings{mustard,
    title = "Towards Multimodal Sarcasm Detection (An \_Obviously\_ Perfect Paper)",
    author = "Castro, Santiago  and
      Hazarika, Devamanyu  and
      P{\'e}rez-Rosas, Ver{\'o}nica  and
      Zimmermann, Roger  and
      Mihalcea, Rada  and
      Poria, Soujanya",
    booktitle = "Proceedings of the 57th Annual Meeting of the Association for Computational Linguistics (Volume 1: Long Papers)",
    month = "7",
    year = "2019",
    address = "Florence, Italy",
    publisher = "Association for Computational Linguistics",
}

@inproceedings{reimers-gurevych-2019-sentence,
    title = "Sentence-{BERT}: Sentence Embeddings using {S}iamese {BERT}-Networks",
    author = "Reimers, Nils  and
      Gurevych, Iryna",
    editor = "Inui, Kentaro  and
      Jiang, Jing  and
      Ng, Vincent  and
      Wan, Xiaojun",
    booktitle = "Proceedings of the 2019 Conference on Empirical Methods in Natural Language Processing and the 9th International Joint Conference on Natural Language Processing (EMNLP-IJCNLP)",
    month = nov,
    year = "2019",
    address = "Hong Kong, China",
    publisher = "Association for Computational Linguistics",
    pages = "3982--3992",
}

@article{MIXCHORD,
      title={On-Policy RL Meets Off-Policy Experts: Harmonizing Supervised Fine-Tuning and Reinforcement Learning via Dynamic Weighting},
      author={Wenhao Zhang and Yuexiang Xie and Yuchang Sun and Yanxi Chen and Guoyin Wang and Yaliang Li and Bolin Ding and Jingren Zhou},
      year={2025},
      journal={arXiv preprint arXiv:2508.11408},
}

@inproceedings{poria-etal-2019-meld,
    title = "{MELD}: A Multimodal Multi-Party Dataset for Emotion Recognition in Conversations",
    author = "Poria, Soujanya  and
      Hazarika, Devamanyu  and
      Majumder, Navonil  and
      Naik, Gautam  and
      Cambria, Erik  and
      Mihalcea, Rada",
    editor = "Korhonen, Anna  and
      Traum, David  and
      M{\`a}rquez, Llu{\'i}s",
    booktitle = "Proceedings of the 57th Annual Meeting of the Association for Computational Linguistics",
    month = jul,
    year = "2019",
    pages = "527--536",
}

@article{nvidia2026cosmos3omnimodalworld,
      title={Cosmos 3: Omnimodal World Models for Physical AI}, 
      author={NVIDIA et al.},
      year={2026},
      journal={arXiv preprint arXiv:2606.02800},
}

@article{wang2025internvl35advancingopensourcemultimodal,
      title={InternVL3.5: Advancing Open-Source Multimodal Models in Versatility, Reasoning, and Efficiency}, 
      author={Weiyun Wang and Zhangwei Gao et al.},
      year={2025},
      journal={arXiv preprint arXiv:2508.18265},
}

@inproceedings{lian2025ov,
  title={OV-MER: Towards Open-Vocabulary Multimodal Emotion Recognition},
  author={Lian, Zheng and Sun, Haiyang and Sun, Licai and Chen, Haoyu and Chen, Lan and Gu, Hao and Wen, Zhuofan and Chen, Shun and Siyuan, Zhang and Yao, Hailiang and others},
  booktitle={Proceedings of the 42nd International Conference on Machine Learning},
  year={2025}
}

@article{bai2025qwen3vltechnicalreport,
      title={Qwen3-VL Technical Report}, 
      author={Shuai Bai and Yuxuan Cai et al.},
      year={2025},
      journal={arXiv preprint arXiv:2511.21631},
}

@inproceedings{10.1145/3706599.3706743,
author = {Ahmadpour, Naseem and Lottridge, Danielle and Fritsch, Jonas and Sas, Corina and Cecchinato, Marta E. and Harrison, Daniel and H{\"o}{\"o}k, Kristina and Foong, Pin Sym and Ijaz, Kiran and Gough, Phillip and Cao, Yidan and Li, Xuefei and Lazem, Shaimaa and Sachathep, Thida},
title = {Affective interaction and affective computing - past, present and future},
year = {2025},
isbn = {9798400713958},
publisher = {Association for Computing Machinery},
address = {New York, NY, USA},
booktitle = {Proceedings of the Extended Abstracts of the CHI Conference on Human Factors in Computing Systems},
articleno = {768},
numpages = {6},
series = {CHI EA '25}
}

%%%%%%%%%%%%%%%%%%%%%%%Appendix%%%%%%%%%%%%%%%%%%%%%%%%%%%%%%%%%%%%%%%%%%%%%%%%%%%%%%%%
\onecolumn
\setcounter{page}{1}
\setcounter{table}{0}
\setcounter{figure}{0}
\setcounter{section}{0}
\setcounter{equation}{0}
\section{Appendix Overview}
In this Appendix, we present the following as an extension to the ones shown in the paper:

\begin{itemize}
    \item Additional Details~(\cref{sup:add_details})
    \item Prompting Protocols~(\cref{sup:prompt})
    %\item Additional Experimental Results (\cref{sup:P2Edetails})
    \item Qualitative Analysis~(\cref{sup:Qualitative analysis})
\end{itemize}

\section{Additional Details}
\label{sup:add_details}
\subsection{Distribution of Data Length}
\begin{figure*}[h]
\centering
\includegraphics[width=1.0\textwidth]{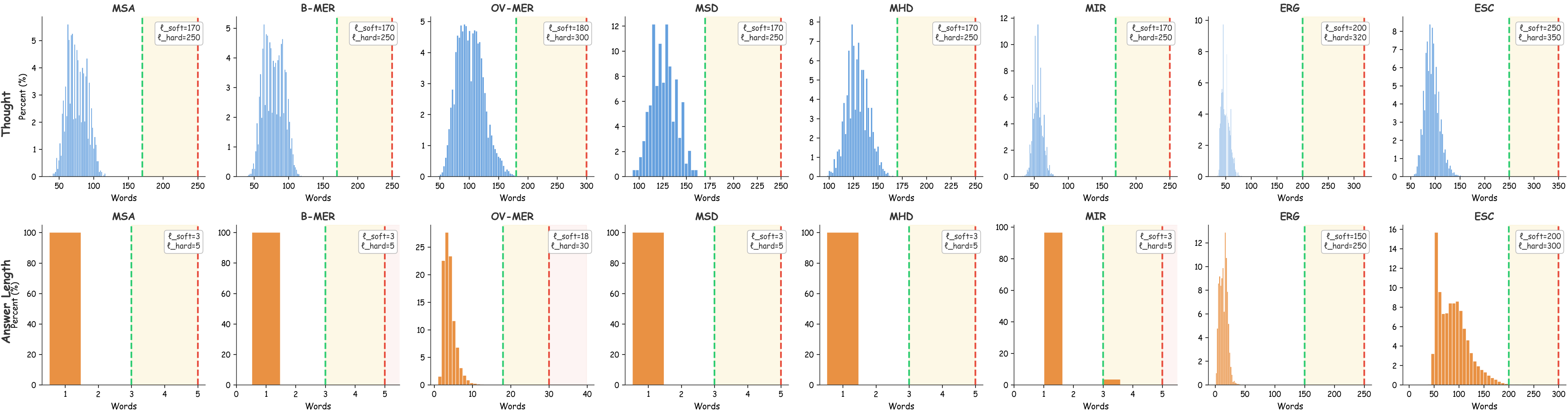}
\caption{\textbf{Length distributions of reasoning trajectories and responses in EmoWorld-130K.} The green and red dashed lines denote the task-specific soft and hard length penalty thresholds, respectively. Notably, the vast majority of samples naturally fall within the penalty-free region, demonstrating that our heuristic thresholds preserve the natural output distribution while exclusively penalizing pathological verbosity.
}
\label{fig:length_distribution}
\end{figure*}

As illustrated in Fig.~\ref{fig:length_distribution}, the natural output lengths of EmoWorld-130K vary significantly across different task paradigms. For instance, perception or understanding tasks (e.g., B-MER, MIR tasks) typically yield concise answers. In contrast, interaction-level tasks (e.g., OV-MER and ESC task) intrinsically necessitate more extensive reasoning trajectories and detailed conversational responses. To accommodate these heterogeneous length biases, we heuristically assign dynamic thresholds grouped by the inherent length characteristics of each task category.
This task-specific thresholding method achieves a vital pragmatic balance: it effectively suppresses pathological verbosity, such as reward hacking or endless rambling during RL exploration, without truncating valid, well-formed reasoning chains. Crucially, as corroborated by Fig.~\ref{fig:length_distribution}, the vast majority of our samples naturally fall within the penalty-free region. This geometric alignment indicates that our heuristic thresholds successfully preserve the natural distribution of the model's output, strictly penalizing only genuinely excessively long generations.
% We do not claim these specific threshold values to be globally optimal; rather, they serve as an empirically robust and pragmatic design choice that stabilizes our multi-task framework. We leave the exploration of systematic grid search or fully adaptive dynamic thresholding mechanisms to future research.

%%%%%%%%%%%%%%%%%%%%%%%%%%%%%%%%sup fig1%%%%%%%%%%%%%%%%%%%%%%%%%%%%%%%%
%%%%%%%%%%%%%%%%%%%%%%%%%%%%%%%%sup fig1%%%%%%%%%%%%%%%%%%%%%%%%%%%%%%%%
%-------------------------------sup tab1---------------------------------
\subsection{Training Data Ratio}

\begin{table}[h]
\centering
\caption{Data mixing proportions (\%) across different training phases.}
\label{tab:data_mixing}
%\resizebox{\columnwidth}{!} % 自动缩放以适应单栏宽度
{
\begin{tabular}{lcccccccc}
\toprule
\textbf{Phase} & \textbf{MSA} & \textbf{MER} & \textbf{OVMER} & \textbf{MIR} & \textbf{MSU} & \textbf{MHU} & \textbf{ERG} & \textbf{ESC} \\
\midrule
\textbf{Curriculum Stage 1} & 7.26 & 9.69 & 58.11 & 10.90 & 0.73 & 3.63 & 9.69 & - \\
\textbf{Curriculum Stage 2} & 3.64 & 5.46 & 45.47 & 8.55 & 0.69 & 3.46 & 18.19 & 14.55 \\
\textbf{Emo-CHord} & 9.43 & 11.31 & 16.50 & 15.32 & 1.29 & 6.58 & 28.28 & 11.31 \\
\bottomrule
\end{tabular}%
}
\end{table}

Table~\ref{tab:data_mixing} details the task sampling proportions across different training phases. Stage 1 prioritizes fundamental perception for robust multimodal grounding, while Stage 2 dynamically amplifies interaction tasks. Finally, \textit{RL mix SFT} introduces ESC and balances the multi-level distribution to foster comprehensive affective synergy without catastrophic forgetting.
%-------------------------------sup tab1---------------------------------
\section{Prompting Protocols for Automated Evaluation and Rubric Scoring}
\label{sup:prompt}
In this section, we provide the complete prompt templates 
in the file \texttt{prompts.py}, which can be found 
at our github code repository. Below we present a condensed version of the key instructions:

\subsection{Perception and Understanding Task Thought Reward Prompt}
\begin{lstlisting}
Instructions:
Use only the provided information. Ignore fluency and formatting unless they affect judgment. Inferring internal states from visual cues is permitted as long as they do not contradict the provided facts.
Task: Evaluate the candidate's reasoning on multimodal tasks (e.g., emotion recognition, intent detection) by distinguishing objective atomic visual behaviors from subjective inferences. Assessment should be based on the extracted behaviors and reasoning content.

1. Visual Fact Consistency
Assess alignment of the model's internal reasoning with provided atomic visual facts. Penalize fabricated observations, invented events, or contradictions. Allow reasonable high-level inferences if not contradicted by observable cues.

Scoring Criteria:
5 points: No claims clearly contradict provided facts or explicit text; no substantial hallucinations.
4 points: Essentially consistent. Minor unsupported details present but do not affect core judgment.
3 points: Partially consistent. Some important inferences are weakly grounded, but overall logic remains acceptable.
2 points: Multiple significant claims lack basis, are exaggerated, or involve selective interpretation of evidence.
1 point: Most reasoning relies on fictional cues or clearly conflicts with evidence.
0 points: No usable basis, missing or irrelevant reasoning, or reasoning explicitly contradicted by evidence.

2. Reasoning-Answer Coherence
Assess internal consistency between reasoning and the final answer (evaluate coherence, not answer correctness). The answer must naturally follow from the reasoning (convergence for single-label; logical support for open-label).

Scoring Criteria:
5 points: Reasoning clearly and directly supports the final answer, with no contradictions or unresolved alternatives.
4 points: Essentially consistent. Answer matches reasoning, with only minor ambiguity or brevity.
3 points: Roughly compatible, but the logical chain from reasoning to answer is incomplete, vague, or omits key clarifying steps.
2 points: Clear mismatch. Some reasoning points in a different direction, or the final answer lacks sufficient support.
1 point: Severe contradiction between reasoning and final answer.
0 points: Missing reasoning or answer, or answer cannot be logically inferred from reasoning.
\end{lstlisting}

\subsection{Interaction Task Thought Reward Prompt}
\begin{lstlisting}
Instruction: Evaluate the reasoning process for empathetic/counseling tasks (ERG/ESC) by comparing candidate reasoning with reference reasoning. Focus on semantic alignment, not wording style.

1. User State Alignment
Criteria: Compare at a semantic level. Focus on capturing the user's situation, emotional state, core conflict, needs, and drivers. Exact wording is not required; equivalent paraphrases are accepted.

Scoring Criteria:
5: Fully aligned. Captures the user's state, emotions, needs, and core conflict with no material drift.
4: Mostly aligned. Minor omissions or wording differences, but the core state is captured.
3: Partially aligned. Some central aspects are captured, but important emotional or situational details are missing or blurred.
2: Weak alignment. Captures only a small part of the user state, or mixes it with notable misunderstanding.
1: Largely misaligned. Describes a substantially different user state or emotional meaning.
0: No usable user-state analysis, or irrelevant to the reference reasoning.

2. Response Strategy Alignment
Criteria: Compare the implied strategy with the reference. Focus on the intended response plan, sequencing, and prioritization (e.g., validation, exploration, reassurance, reframing).

Scoring Criteria:
5: Fully aligned. Matches the reference strategy, support goal, and prioritization of moves.
4: Mostly aligned. Same broad strategy with only minor omissions or reduced specificity.
3: Partially aligned. Relevant but misses an important component of the reference strategy.
2: Weak alignment. Only superficial overlap or a noticeably off-target direction.
1: Strong mismatch. Substantially different or counterproductive strategy.
0: No usable strategy, or irrelevant to the reference reasoning.
General Instructions:
\end{lstlisting}

\subsection{Interaction Task Evaluation Prompt}
\begin{lstlisting}
##ERG task:
Instruction: Evaluate the response on a scale of 1 (Poor) to 5 (Excellent) across the following dimensions.

Dimension 1: Empathy

Score Rubric:
1: Completely misses or contradicts the seeker's emotional state; response is cold, dismissive, or overtly inappropriate.
2: Shows minimal recognition of emotion; response is generic (e.g., "I'm sorry to hear that") with no evidence of true understanding.
3: Acknowledges the emotion at a surface level; response is adequate but lacks depth, specificity, or genuine warmth.
4: Demonstrates clear understanding of the emotional state; response is warm, validating, and reasonably tailored to the seeker's situation.
5: Deeply and precisely captures the nuanced emotional state; response feels genuinely human, highly validating, and emotionally resonant.

Dimension 2: Coherence & Consistency

Score Rubric:
1: Completely incoherent, contradictory, or irrelevant to the seeker's message.
2: Mostly inconsistent or contains noticeable logical gaps; relevance is weak.
3: Generally coherent and relevant, but contains minor inconsistencies, vague connections, or slight drift.
4: Consistent and well-connected to the context; logical flow is clear with only trivial issues.
5: Perfectly coherent, logically airtight, and seamlessly aligned with the seeker's context and intent.

Dimension 3: Informativeness

Score Rubric:
1: Purely templatized or meaningless filler; adds zero informational or conversational value.
2: Mostly generic platitudes with negligible substantive content; could apply to any situation.
3: Contains some meaningful content but still relies noticeably on generic or templatized language.
4: Substantive and reasonably specific; provides genuine value (insight, relevant question, useful framing) with minimal templatization.
5: Highly informative and uniquely tailored; offers profound insight, a precisely targeted question, or genuinely useful perspective with no detectable templatization.
\end{lstlisting}

\begin{lstlisting}
##ESC task:
Instruction: Evaluate the response on a scale of 1 (Poor) to 5 (Excellent) across the following dimensions.

Dimension 1: Empathy

Score Rubric:
1: Misses, dismisses, or contradicts the patient's feelings; cold or invalidating.
2: Minimal acknowledgment of emotion; generic or perfunctory.
3: Recognizes the general feeling but remains surface-level.
4: Clearly understands and validates the patient's emotional state with warmth and presence.
5: Precisely captures the emotional nuance and feels deeply human, attuned, and validating.

Dimension 2: Support Skill

Score Rubric:
1: No discernible support strategy, or actively harmful technique (e.g., toxic positivity, blame, excessive advice).
2: Attempts a strategy but executes poorly (e.g., closed yes/no question, irrelevant self-disclosure, shallow reassurance).
3: Uses one or more recognizable support strategies correctly, but selection or timing is suboptimal (e.g., advice before validation).
4: Deploys appropriate strategies with good timing and execution; demonstrates competent conversational support skills.
5: Masterful, context-sensitive use of multiple support strategies with impeccable timing; reflects professional-level support skill.

Dimension 3: Overall Effectiveness

Score Rubric:
1: Likely to leave the seeker feeling worse, unheard, or frustrated; response is counter-therapeutic.
2: Minimal positive impact; seeker would likely feel little change or mild disappointment.
3: Moderately helpful; seeker receives some comfort or direction, but the response lacks lasting impact or full trust.
4: Clearly helpful; seeker would likely feel noticeably heard, validated, and somewhat emotionally relieved.
5: Exceptionally effective; seeker would likely feel deeply understood, genuinely comforted, and measurably better equipped to cope.
\end{lstlisting}

\section{Qualitative Analysis}
\label{sup:Qualitative analysis}
%%%%%%%%%%%%%%%%%%%%%%%%%%%%%%%%sup fig2%%%%%%%%%%%%%%%%%%%%%%%%%%%%%%%%
\begin{figure*}[h]
\centering
\includegraphics[width=1.0\textwidth]{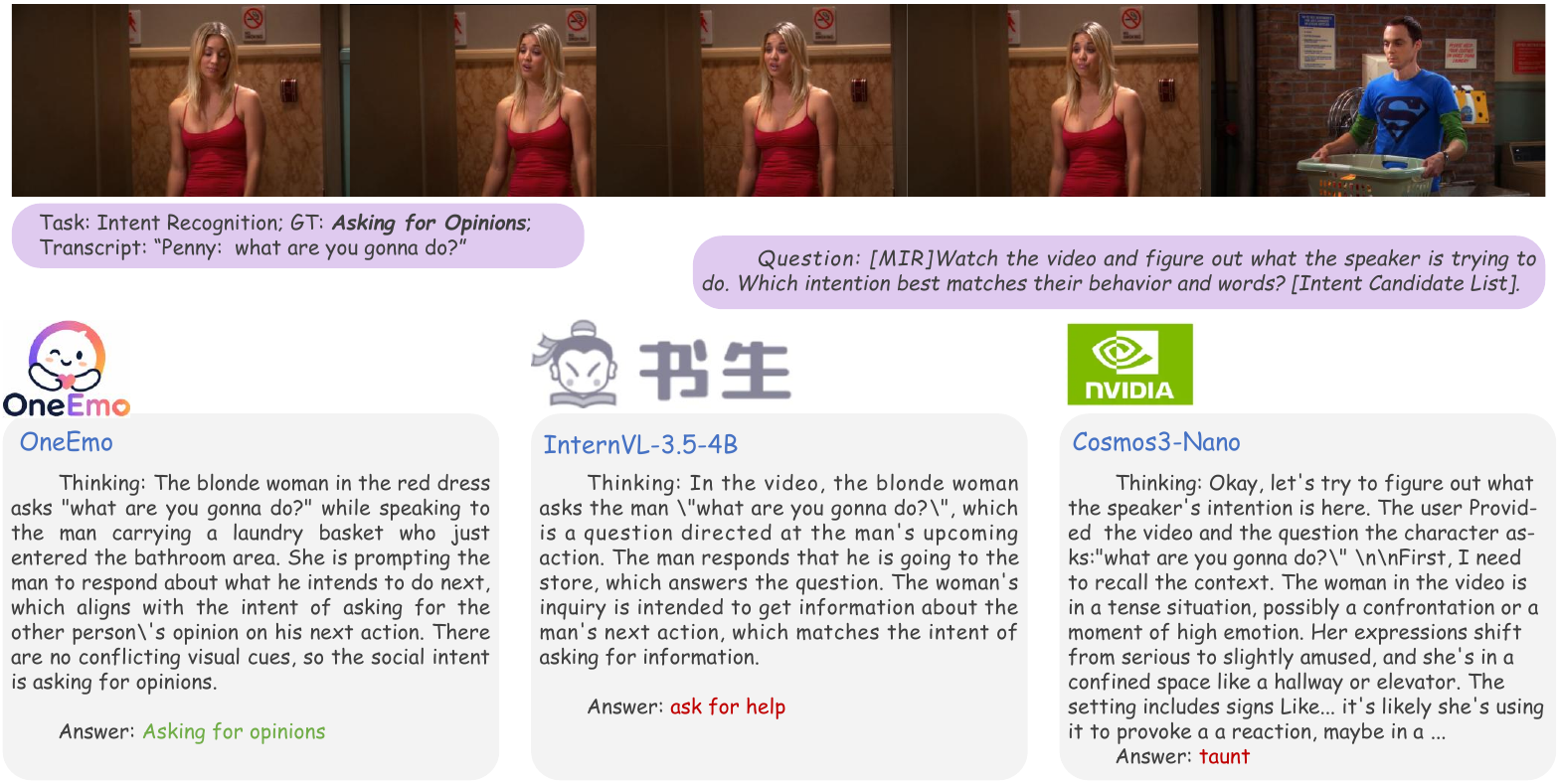}
\caption{Qualitative comparison of reasoning trajectories on a multimodal intent recognition task. While OneEmo accurately grounds both visual and textual cues to deduce the correct intent, the baseline models exhibit label mapping errors (InternVL-3.5-4B) and severe contextual hallucinations (Cosmos3-Nano).
}
\label{fig:qualitative_analysis}
\end{figure*}
%%%%%%%%%%%%%%%%%%%%%%%%%%%%%%%%sup fig2%%%%%%%%%%%%%%%%%%%%%%%%%%%%%%%%
%%%%%%%%%%%%%%%%%%%%%%%%%%%%%%%%%%%%%%%%%%%%%%%%%%%%%%%%%%%%%%%%%%%%%%%%%%%%%%%
Fig.~\ref{fig:qualitative_analysis} compares reasoning trajectories for intent recognition. OneEmo accurately extracts multimodal cues (e.g., characters and objects) to logically deduce the correct intent. In contrast, InternVL-3.5-4B correctly interprets the semantic inquiry but fails in label mapping. Meanwhile, Cosmos3-Nano exhibits severe contextual hallucinations, fabricating a tense confrontation in an elevator, resulting in an erroneous taunt prediction. 

%%%%%%%%%%%%%%%%%%%%%%%%%%%%%%%%%%%%%%%%%%%%%%%%%%%%%%%%%%%%%%%%%%%%%%%%%%%%%%%


% \newpage

 

% \end{IEEEbiography}



% \vfill

\end{document}